\global\def\draftcontrol{0}

   \def\versionno{ stability }

\catcode`\@=11

\expandafter\ifx\csname draftcontrol\endcsname\relax\global\def\draftcontrol{0} 
\fi 

{\count255=\time\divide\count255 by 60 
\xdef\hourmin{\number\count255} 
\multiply\count255 by-60\advance\count255 by\time 
\xdef\hourmin{\hourmin:\ifnum\count255<10 0\fi\the\count255}} 
\def\draftdate{\number\month/\number\day/\number\year\ \ \ \hourmin } 

\newcommand\makepapertitle{\par

  \begingroup 
    \renewcommand\thefootnote{\@fnsymbol\c@footnote}%
    \def\@makefnmark{\rlap{\@textsuperscript{\normalfont\@thefnmark}}}%
    \long\def\@makefntext##1{\parindent 1em\noindent 
            \hb@xt@1.8em{%
                \hss\@textsuperscript{\normalfont\@thefnmark}}##1}%
     \newpage 
     \global\@topnum\z@   
     \@makepapertitle 
     \thispagestyle{empty}\@thanks 
  \endgroup 
  \setcounter{footnote}{0}%
  \global\let\thanks\relax 
  \global\let\makepapertitle\relax 
  \global\let\@makepapertitle\relax 
  \global\let\@thanks\@empty 
  \global\let\@author\@empty 
  \global\let\@date\@empty 
  \global\let\@title\@empty 
  \global\let\title\relax 
  \global\let\author\relax 
  \global\let\date\relax 
  \global\let\and\relax 
  \def\version{\let\version\@version\@gobble} 
} 
\def\@makepapertitle{%
  \newpage 
   \ifnum\draftcontrol=1 {} 
   \version\versionno 
   \vskip 5em%
   \else 
   \hfill\hbox to 3cm {\parbox{4cm}{\@pubnum}\hss}%
   \vskip 5em%
   \fi 
   \begin{center}%
   \let \footnote \thanks 
      {\hskip -0\textwidth \hbox to 1\textwidth%
        {\centerline{\Large\bf{\noindent\@title}}}}%
     \vskip 2em%
     {\normalsize
       \lineskip .5em%
       \begin{tabular}[t]{c}%
         \@author 
       \end{tabular}\par}%
     \vskip 1em%
     {\@bstract}%
     \end{center}%
     \vfill
     \@date%
     \vskip 1.5em%
   \par 
} 

\gdef\@pubnum{} 
\def\pubnum#1{%
  \gdef\@pubnum{#1}} 

\gdef\@bstract{} 
\def\Abstract#1{%
  \gdef\@bstract{%
   \parbox{\textwidth-0pc}{%
   \centerline{\bf Abstract}\penalty1000 
   \noindent
   \renewcommand\baselinestretch{1.0} 
   {#1}}} 
} 

\gdef\@email{}
\def\email#1{%
   \gdef\@email{%
   Email: {\tt #1}}
}

\def\ps@paper{\let\@mkboth\@gobbletwo%
     \ifnum\draftcontrol=1 
        \def\@oddfoot{\hbox to \textwidth{\tiny \versionno \hfil\tiny\draftdate}%
        \hskip -\textwidth \hbox to \textwidth{\hfil\rm\thepage\hfil}}%
     \else\def\@oddfoot{\hbox to \textwidth{\hfil\rm\thepage\hfil}} 
     \fi 
     \let\@evenfoot\@oddfoot 
} 

\def\body{\clearpage 
          \pagestyle{paper} 
        } 
\newenvironment{acknowledgments}{%
\vskip 3.25ex 
\noindent {\bf Acknowledgments} 
} 

\def\@version#1{\ifnum\draftcontrol=1 
\typeout{}\typeout{#1}\typeout{} 
\vskip3mm\centerline{\hbox{\fbox{\normalsize{\tt DRAFT -- #1 -- } 
                   {\draftdate}}}}\vskip3mm 
\fi} 
\let\version\@version 
\long\def\eqlabel#1{\ifnum\draftcontrol=1 
                    \tag@false  
                    \tag*{(\theequation) \hbox to -0.2cm{\hspace{0cm}\small{#1}\hss}} 
                    \refstepcounter{equation}  
                    \edef\@currentlabel{\theequation} 
                    \ltx@label{#1}          
                    \else 
                    \label{#1} 
                    \fi 
                    } 
\let\st@bibitem\@bibitem 
\let\st@lbibitem\@lbibitem 
\ifnum\draftcontrol=1 
  \def\@bibitem#1{%
    \st@bibitem{#1}\a@@label{#1}\ignorespaces} 
  \def\@lbibitem[#1]#2{%
    \st@lbibitem[#1]{#2}\a@@label{#2}\ignorespaces} 
  \def\a@@label#1{%
    \gdef\a@lab{\smash{\normalfont\small#1}} 
    \ifvmode 
      \if@inlabel 
        \global\setbox\@labels\hbox{%
          \llap{\a@lab\let\a@lab\relax 
                \kern\@totalleftmargin\kern\marginparsep}%
          \box\@labels}%
      \fi 
    \fi} 
\fi 

\documentclass[12pt,letterpaper]{article} 

\usepackage[ps,dvips,matrix,arrow,frame,import,curve,color]{xy}
\usepackage{amsmath,bm,amsfonts,amssymb,array,calc,amsthm,rotating}
\usepackage[nosort]{cite} 
\usepackage{graphicx}

\renewcommand\baselinestretch{1.25} 
\renewcommand\section{\@startsection {section}{1}{\z@}%
                                   {-3.5ex \@plus -1ex \@minus -.2ex}%
                                   {2.3ex \@plus.2ex}%
                                   {\normalfont\large\bfseries}} 
\renewcommand\subsection{\@startsection{subsection}{2}{\z@}%
                                   {-3.25ex\@plus -1ex \@minus -.2ex}%
                                   {1.5ex \@plus .2ex}%
                                   {\normalfont\normalsize\bfseries}} 
\renewcommand\subsubsection{\@startsection{subsubsection}{3}{\z@}%
                                   {-3.25ex\@plus -1ex \@minus -.2ex}%
                                   {1.5ex \@plus .2ex}%
                                   {\normalfont\normalsize\it}} 
\renewcommand\paragraph{\@startsection{paragraph}{4}{\z@}%
                                   {-3.25ex\@plus -1ex \@minus -.2ex}%
                                   {1.5ex \@plus .2ex}%
                                   {\centering\normalfont\normalsize\it}} 
\renewcommand\subparagraph{\@startsection{subparagraph}{5}{\z@}%
                                   {-1.25ex\@plus -1ex \@minus -.2ex}%
                                   {0ex \@plus .2ex}%
                                   {\normalfont\normalsize\it}} 
\newcommand\heading[1]{\paragraph{#1}}

\numberwithin{equation}{section}

\long\def\@makecaption#1#2{%
  \vskip\abovecaptionskip
  \sbox\@tempboxa{{\small \bf #1:} {\small #2}}%
  \ifdim \wd\@tempboxa >\hsize
    {\small\bf #1:} {\small #2}\par
  \else
    \global \@minipagefalse
    \hb@xt@\hsize{\hfil\box\@tempboxa\hfil}%
  \fi
  \vskip\belowcaptionskip}

\renewcommand*\l@section[2]{%
  \ifnum \c@tocdepth >\z@
    \addpenalty\@secpenalty
    \addvspace{.5em \@plus\p@}%
    \setlength\@tempdima{1.5em}%
    \begingroup
      \parindent \z@ \rightskip \@pnumwidth
      \parfillskip -\@pnumwidth
      \leavevmode \bfseries
      \advance\leftskip\@tempdima
      \hskip -\leftskip
      #1\nobreak\hfil \nobreak\hb@xt@\@pnumwidth{\hss #2}\par
    \endgroup
  \fi}
\renewcommand*\l@subsection{\addvspace{.0em \@plus\p@}\@dottedtocline{2}{1.5em}{2.3em}}
\renewcommand*\l@subsubsection{\addvspace{-.2em \@plus\p@}\@dottedtocline{3}{3.8em}{3.2em}}

\def\revise#1       {\raisebox{-0em}{\rule{3pt}{1em}}%
                     \marginpar{\raisebox{.5em}{\vrule width3pt\ 
                     \vrule width0pt height 0pt depth0.5em 
                     \hbox to 0cm{\hspace{0cm}{%
                     \parbox[t]{4em}{\raggedright\footnotesize{#1}}}\hss}}}}

\def\zet          {{\mathbb Z}} 

\def\del          {\partial} 
\def\delbar       {\bar\partial}

\def\tr           {{\rm Tr}}

\def\sqr#1#2{{\vcenter{\vbox{\hrule height.#2pt   
 \hbox{\vrule width.#2pt height#1pt \kern#1pt 
 \vrule width.#2pt}\hrule height.#2pt}}}}

\newcommand\ba{\begin{eqnarray}}
\newcommand\ea{\end{eqnarray}}

\def\Dslash{\,\,{\raise.15ex\hbox{/}\mkern-12mu D}}
\def\Dbarslash{\,\,{\raise.15ex\hbox{/}\mkern-12mu {\bar D}}}
\def\delslash{\,\,{\raise.15ex\hbox{/}\mkern-9mu \partial}}
\def\delbarslash{\,\,{\raise.15ex\hbox{/}\mkern-9mu {\bar\partial}}}
\def\pslash{\,\,{\raise.15ex\hbox{/}\mkern-9mu p}}
\def\calDslash{\,\,{\raise.15ex\hbox{/}\mkern-12mu {\cal D}}}

\newcommand{\hh}{{1\over 2}}

\renewcommand{\ll}{_}
\newcommand{\uu}{^}
\newcommand{\pp}{\partial}
\renewcommand{\L}{\Lambda}
\renewcommand{\exp}[1]{{\rm exp}\{#1\}}

\newcommand{\m}{\mu}

\renewcommand{\dag}{{}^\dagger{}}

\renewcommand{\m}{\mu}
\newcommand{\n}{\nu}
\newcommand{\s}{\sigma}
\renewcommand{\t}{\tau}
\newcommand{\G}{\Gamma}
\newcommand{\g}{\gamma}
\renewcommand{\a}{\alpha}
\renewcommand{\r}{\rho}

\renewcommand{\o}{\omega}
\newcommand{\e}{\epsilon}
\renewcommand{\O}{\phi}

\newcommand{\sqd}{^2}
\newcommand{\zb}{{\bar{z}}}

\renewcommand{\hh}{{1\over 2}}

\newcommand{\eee}[1]{\ba{#1}\ea}
\renewcommand{\th}{\alpha}
\renewcommand{\t}{\tau}

\renewcommand{\b}{\beta}

\newcommand{\llsk}{\hskip .5in}

\newcommand{\st}{{}^*}

\newcommand{\pr}{^\prime {}}

\newcommand{\apr}{{\alpha^\prime} {}}

\newcommand{\IZ}{\relax\ifmmode\mathchoice
{\hbox{\cmss Z\kern-.4em Z}}{\hbox{\cmss Z\kern-.4em Z}}
{\lower.9pt\hbox{\cmsss Z\kern-.4em Z}} {\lower1.2pt\hbox{\cmsss
Z\kern-.4em Z}}\else{\cmss Z\kern-.4em Z}\fi} \font\cmss=cmss10
\font\cmsss=cmss10 at 7pt
\newcommand{\inbar}{\,\vrule height1.5ex width.4pt depth0pt}
\newcommand{\IC}{{\relax\hbox{$\inbar\kern-.3em{\rm C}$}}}
\newcommand{\IQ}{{\relax\hbox{$\inbar\kern-.3em{\rm Q}$}}}
\newcommand{\IP}{\relax{\rm I\kern-.18em P}}

\newcommand{\ed}{\dot{e}}

\renewcommand{\l}{\lambda}

\newcommand{\htt}{\tilde{h}}

\newcommand{\lt}{\tilde{L}}
\newcommand{\mflw}{(-1)^{F_{L_W}}}

\newcommand{\ff}{{1\over 4}}

\renewcommand{\o}{\omega}

\newcommand{\ct}{\tilde{c}}

\renewcommand{\L}{\Lambda}
\renewcommand{\pr}{{}^\prime{}}

\newcommand{\at}{\tilde{\alpha}}

\newcommand{\pst}{\tilde{\psi}}

\renewcommand{\at}{{\tilde{\alpha}}}

\newcommand{\IR}{\relax{\rm I\kern-.18em R}}
\def\blfootnote{\xdef\@thefnmark{}\@footnotetext}

\renewcommand{\bm}{\begin{matrix}}
\renewcommand{\em}{\end{matrix}}

\newcommand{\lno}{\left .}
\newcommand{\rno}{\right .}

\newcommand{\rba}{\right |}

\newcommand{\up}[1]{^{({#1})}{}}
\newcommand{\upb}[1]{^{<{#1}>}{}}

\renewcommand{\tr}{{\rm tr}}

\newcommand{\bbb}{\ba\begin{array}{c}}
\renewcommand{\eee}{\label{}\end{array}\ea}
\newcommand{\een}[1]{\label{#1}\end{array}\ea}

\def\hilo{{}_{{}_{{}_{{}_{{}_{}}}}} {}^{{}^{{}^{}}}}

\def\lrdd{\left ( ~}
\def\rrdd{\hilo \right )}
\def\lsqq{\left [ ~}
\def\rsqq{\hilo \right ]}

\newcommand{\kket}[1]{\left | {#1} \right \rangle }

\def\bi{\begin{itemize}}
\def\ei{\end{itemize}}

\def\qb{\bar{q}}

\renewcommand{\htt}{{\tilde{h}}}

\def\ed{\end{document}}

\def\tb{\bar{\tau}}

\def\nsm{{\rm NS}_-}
\def\nsp{{\rm NS}_+}

\def\mfls{(-1)^{F_{L_S}}}
\def\mfrw{(-1)^{F_{R_W}}}

\def\btt{\begin{table}}
\def\ett{\end{table}}
\def\bta{\begin{tabular}}
\def\etaa{\end{tabular}}

\def\lt{{\tilde{\lambda}}}

\def\rrp{{\rm R}_+}

\def\rri{{\rm R}_i}

\def\rpmi{{\rm R}_{\pm i}}

\def\nsi{{\rm NS}_i}

\def\mot{- {1\over \tau}}
\def\vsk{\vskip 0.5in}
\def\boi{{\bf I}}
\def\phum{\phi^{-1}}

\def\bfft{\tilde{{\bf F}}}
\def\fft{{\tilde{F}}}

\def\htt{{\tilde{h}}}
\def\ct{{C_T}}
\def\wh{{\hat{w}}}
\def\xf{x\ll{\rm{\bf{F}}}}
\def\nh{{\hat{n}}}
\def\Ch{\hat{C}}

\def\nf{n_f}
\def\Nf{N_f}
\def\Nft{\tilde{N}_f}
\def\wf{w_f}
\def\Nt{\tilde{N}}
\begin{document}


\title{Worldsheet CFTs for Flat Monodrofolds}

\pubnum{%
hep-th/0604191}
\date{April 2006}

\author{
Simeon Hellerman$^{1}$ 
and Johannes Walcher$^{2}$ 
\\[0.4cm]
\it School of Natural Sciences, Institute for Advanced Study\\
\it Princeton, New Jersey, USA\\
$^{1}${\tt simeon@ias.edu}
, $^{2}${\tt walcher@ias.edu}
\\[.5cm]
}

\Abstract{
We resolve a puzzle in the theory of strings propagating 
on locally flat spacetimes with nontrivial Wilson lines
for stringy $\IZ\ll N$ gauge symmetries. We find that
strings probing such backgrounds are described by consistent
worldsheet CFTs. The level mismatch in the twisted sectors is
compensated by adjusting the quantization of momentum of
strings winding around the Wilson line direction in units of 
$1/RN^2$ rather than $1/RN$, as might have been classically 
expected. We demonstrate in various examples how this improvement
of the naive orbifold prescription leads to satisfaction of 
general physical principles such as level matching and closure 
of the OPE. Applying our techniques to construct a Wilson line 
for T-duality of a torus in the type II string (``T-fold''), we 
find a new 7D solution with ${\cal N} = 1$ SUSY where the moduli 
of the fiber torus are fixed.  When the size of the base becomes 
small this simple monodrofold exhibits enhanced gauge symmetry 
and a self-T-duality on the $S\uu 1$ base. 
}

\makepapertitle

\body

\version\versionno

\vskip 1em

\setcounter{tocdepth}{2}
\tableofcontents
\newpage

\section{Introduction}
\label{intro}

Worldsheet conformal field theories (CFTs) based on free fields 
represent the simplest type of weakly coupled string theory.  
The simplest of all are the cases in which 
the target space has no invariants which distinguish
it from flat space.  This set of examples includes
Narain compactification, but is not limited to it.  For
instance, the Klein bottle can be thought of
as an $S\uu 1\ll{fiber}$ fibered over $S\uu 1\ll{base}$,
with an identification of the fiber under reflection
as one traverses the base.  Following particle physics
terminology, we will refer to such parallel transport, which
is locally trivial but nontrivial around noncontractable
cycles in the base, as a \it Wilson line. \rm  In string
theory, such constructions have been referred to
as \it T-folds, monodrofolds, \rm or when the
fiber and base are tori, as \it twisted tori. \rm
\cite{HMW,hull1,FWW,hull2,STW,FW,SHsnew,hellerman,hull3}

It is widely believed that any exact symmetry of string
theory must be a \it gauge \rm rather than a \it global \rm
symmetry, including discrete symmetries not embedded
in any continuous group.  The defining property of
a gauge symmetry $g$ is that it is a \it local redundancy
of description, \rm which is to say, it should be
possible to parallel transport an object around a closed
loop and find that it comes back to itself up to a
transformation by $g$.  In the case that the loop is
noncontractable, there are no local observables which
can distinguish such a parallel transport law from the
trivial one, and hence no possible nonzero
contributions to a tree-level energy for such a configuration.
\footnote{
Quantum effects such as Casimir forces can
generate nonzero energies for Wilson line backgrounds,
but we will focus in this paper on tree-level physics.}
{} From this we conclude the following:
\begin{center}
\framebox{
\parbox{\textwidth-.5in}{
\scshape Given any string theory ${\cal W}$ in $\IR\uu D$ 
with an unbroken
internal symmetry $g$, there exists a Wilson line background
of ${\cal W}$ on ${\IR\uu {D-1}}\times S\uu 1$ in which
states are identified with themselves up to $g$ when
parallel transported around the $S\uu 1$. \rm}}
\end{center}
If the original string theory ${\cal W}$ has a weak coupling 
expansion, and we consider a symmetry $g$ which
commutes with the limit $g\ll s\to 0$, then the resulting
Wilson line theory should also have a weak coupling expansion
characterized by a fundamental string worldsheet CFT.

When the ``fiber'' is a torus (namely, the internal worldsheet
theory ${\cal I}$
of ${\cal W}$ is a free CFT), we can be even more precise about 
the properties such a solution should have.  The $\apr$ 
expansion of the $\b$-function equations is a derivative 
expansion in the target space. There are no invariants in 
any patch of space which differentiate between the
Wilson line background and flat space.  
It follows that any Wilson line background must be represented 
by a worldsheet CFT which locally in target space is a
free CFT.  

To learn more about the nature of this free CFT,
observe that the Wilson line background is in some
sense a quotient of a trivial background
of ${\cal W}$ on $\IR\uu{D-1}\times S\uu{1\pr}$, where
$S\uu{1\pr}$ is a covering circle. Fix the size of the circle 
to be $2\pi R\ll{base}$, parametrized by the coordinate $X$.
To the extent that string degrees of freedom are
(approximately) local fields in $X$ space, we should be able to 
describe the Wilson line for a $\IZ\ll N$ symmetry
as a quotient of an $N$-times bigger covering circle with radius
$R\pr = N R\ll{base}$
by a $\IZ\ll N$ which acts as a shift $g\ll X: X\to 
X + 2\pi R\ll{base}$
by one $N\uu{\underline{\rm{th}}}$ the size of the 
covering circle, combined with the action of $g$ on all fields.

There is an argument---not quite precise---suggesting such a 
quotient should be accessible through the famous \it orbifold 
\rm construction \cite{DHVW}. Certainly the projection
onto gauge-invariant states is the correct one
for strings which are approximately local fields in 
$X$ rather than extended objects.  That is to say, starting
from ${\mathcal I}\times S^{1\pr}$ and imposing invariance 
under $g\times g\ll X$ projects onto states whose $X$-dependence 
is $\exp{i (n + \th) X / R\ll{base}}$
for some $n$ and whose phase under $g$
is $\exp{- 2\pi i \th}$.  So (approximately) local fields
get parallel transported around the circle with a
phase depending on their $g$-transformations.  This is
precisely the definition of a Wilson line.

In the twisted sectors, however, the naive orbifold construction
of the Wilson line background can begin to
break down.  If $g$ acts asymmetrically on
left- and right-moving degrees of freedom, sectors twisted
by $g$ can have a level mismatch $L\ll 0 - \tilde{L}\ll 0$
which cannot in general be removed by acting with
oscillators on the twisted ground states.
We will give examples of the breakdown in the next
section, but suffice it to say that the failure
of level matching is a true, nonempty constraint on
the orbifold construction: not every exact
symmetry $g$ gives rise to a consistent, modular
invariant theory through the orbifold construction.

Combining $g$ with $g\ll X$ does not cure the
problem (see \cite{FWW}).  That is to say, orbifolding 
${\mathcal I}\times S^{1\pr}$ by $g\times g\ll X$ yields 
a modular invariant theory if and only if orbifolding by $g$
alone does so.  This presents a paradox, because
the argument for the existence of a Wilson line background
is based on what should be an ironclad principle
in string theory: 
\begin{center}
\framebox{
\parbox{\textwidth-.5in}{\scshape Consistent boundary conditions 
for consistent string theories should yield new consistent string 
theories. \rm}}
\end{center}
We refer to this as the 'consistent + consistent = consistent'
principle, or CCC principle for short.  

We will see in this paper that the CCC principle is in fact vindicated 
on the string worldsheet. Namely, we will show that there always exists 
a consistent choice of momentum fractionation in the twisted sectors 
such that level matching, modular invariance and closure of the OPE can 
be maintained. That is, by allowing the momentum in the base to have 
certain values not necessarily satisfying $N p\ll {base} R\ll{base} 
\in \IZ$, we will restore level matching and thereby save modular 
invariance.  This works because the twisted sectors also carry 
winding in the $X$ direction, so the nonzero momentum and winding 
contribute to the level mismatch $\tilde{L}\ll 0 - L \ll 0$ by 
an amount $- p\ll{base} R\ll{base} w\ll{base}$ which precisely 
cancels the level mismatch due to asymmetric Casimir energies. 
This is in disctinction to a standard $\IZ_N$ orbifold construction,
in which level mismatch is restricted to be in $\zet/N$ and can 
be canceled by adding oscillator energy. The important point is
that both when the level mismatch is in $\zet/N$ and when it is
not (so the standard orbifold prescription will apply), we can
construct a Wilson line background for the symmetry, $g$. We
will refer to this procedure as ``wilsonization'', and distinguish
the ``tame'' case which does not require momentum fractionation beyond
$1/R_{base}N$ from the more general, ``wild'', case.

It should also be noted that even when momenta are fractionated 
in units of $1/RN^2$ in the
twisted sectors, in any given twisted sector, the base momenta differ 
only by multiples of $1/RN$. In this way, it is possible to reconcile
the additional momentum fractionation with the closure of the OPE.

The presence of fractional momentum with $p\ll{base}$
not in $\IZ / N R$
would seem to mean that the modes carrying such momentum are 
non-single-valued on the covering space. However this is not 
quite the right conclusion.  These modes also carry winding, 
and so for large $R$ we should think of them as long winding 
strings with excitations on them carrying momentum along the
direction of the string.  In fact we shall see that the 
consistent treatment of the effective theory on the long 
string not only allows but demands the inclusion of sectors 
with $p\ll {base}$ not in $\IZ / N R$.

We also have to wonder how this momentum fractionation can be 
reconciled with the naive expectation that the Wilson line theories
should be accessible by orbifolding. As we will show, one can indeed
reformulate our ``wilsonization of $g$ over $S^1$'' as an 
orbifoldization. What this requires is to let the $\IZ_N$ action depend
by a phase on the winding around the covering circle $S^{1\pr}$ 
(which is winding $\bmod N$ around the base circle $S^1$). In other 
words, the orbifold action on ${\mathcal I}\times S^{1\pr}$ does 
not factorize between base and fiber. 
The inclusion of these extra phases in the orbifold action is somewhat 
reminiscent of discrete torsion (by viewing
winding around $S^{1\pr}$ as the twist quantum number of yet another 
$\IZ_N$ orbifold). But it should be remembered that discrete torsion 
in orbifolds is usually a relative and not an absolute concept 
whereas in the present case we do not have a choice in picking 
these phases. In this sense, the wilsonization technique we 
will introduce in this paper can be viewed either as a 
refinement or as a generalization of the standard orbifold 
construction. 

The outline of the paper is as follows. We will begin in the next 
section by studying some rather simple examples in the bosonic 
string, taken from both the ``tame'' and ``wild'' category. 
%
%
%
The first of those examples involves 32 free nonchiral
fermions as the fiber
theory and various discrete symmetries in $Spin(32)\ll L$
as the symmetry implemented by the Wilson line. The
second example is a Wilson line for T-duality of a single circle
at the self-dual radius. We will also study an example
in which the fiber theory is not built on free fermions or bosons, 
but consists of an $SU(2)$ WZW model. This serves to illustrate that
our methods are applicable also for fibering more general 
rational conformal field theories. 

In section 3, we will then outline a general framework in which
to understand wilsonization as a new worldsheet construction. This
method is then the worldsheet realization of the CCC principle,
and we hope that if the reader's confidence
in the CCC principle was ever 
unsettled, before the end of section 3 it will be completely restored.

Once the wilsonization technique is firmly established, we will then 
more leisurely present an application to the superstring
in section 4. Specifically, 
we will construct a 7-dimensional background of the type II 
superstring with ${\cal N}=1$ supersymmetry which is not equivalent 
to any geometric string theory. 
This example implemements the transformation $\t\to - {1\over \t}, 
\r\to - {1\over \r}$ on a $T\uu 2$ fiber as a Wilson line around 
an $S\uu 1$ base. Sixteen supercharges are preserved by the Wilson 
line, which acts nontrivially only on the gravitini coming from 
left-moving spin fields on the string.

This example is interesting in itself, mainly because of its
scarcity of massless fields. But we wish to reemphasize that the
chief purpose of this paper is to show that the CCC principle is
beyond any doubt satisfied in string theory and is useful for the
construction of new, in general non-geometric, string backgrounds.
In simple situations this can be verified explicitly at 
the worldsheet level. 

We conclude the paper in section 5.

\section{Wilson line theories in the bosonic string}\label{bosonic}

Before laying out a general theory
of string theory Wilson lines, let us
study some examples to get the general idea.
The bosonic string is the simplest arena in which
to construct Wilson line solutions both of the wild
and tame kind.

\subsection{Chiral $\zet\ll 2$ symmetries
in a free fermion theory}

Consider the bosonic string on $\IR\uu{8,1} \times S\uu 1
\times {\cal I}$, where ${\cal I}$ is an internal
theory of $c = 16$.  We will take the internal
theory ${\cal I}$ to consist of an $SO(32)\ll L
\times SO(32)\ll R$ current algebra ${\cal C}$ at level one,
realized as $32$ free fermions
of each chirality.  We will let the $(-1)\uu {F\ll W}$
projection on fermions be the maximally chiral one -- that is,
we restrict to states with left- and right-moving worldsheet
fermion number separately even: $\mfrw = \mflw = +1$, and
allow right- and left-moving SO(32) spinor sectors independently.

Given our fiber theory ${\cal I}$, we wish to consider Wilson
lines around $x\ll 9$ for discrete symmetries $g$ of ${\cal I}$.
For simplicity we will always consider $g$ to be $\IZ\ll 2$ in
its action on fermions.  (This will sometimes mean that it is $\IZ\ll 4$
in its action on SO(32) spinor states.)
\footnote{A sector with
periodic fermions is often
called a 'Ramond'
boundary condition for the fermions, but to avoid confusion
of terminology we shall only use the term 'Ramond' when
we study the superstring, to
describe the boundary conditions of the worldsheet supercurrents.}

Let us
now analyze some of the simplest examples.  We can
classify the possible actions by the numbers $k\ll L, k\ll R$ of
worldsheet fermions which are odd under the Wilson
line symmetry $g$.  To simplify the
problem further we consider the case $k\ll R = 0$ where the
action on all right-moving chiral operators is trivial.

The exact symmetry group of the theory is 
$Spin(32)\ll L / \IZ\ll 2 \times Spin(32)\ll R / \IZ\ll 2$.

\subsubsection{The case $k\ll L = 0$}

The simplest $\IZ\ll 2$ symmetry which is purely chiral is
$-1$ to the total left-moving SO(32) spinor number. 
This is the quantum symmetry which acts with a $-1$ on
sectors with periodic left-moving fermions.  This 
corresponds to $k\ll L = 0$, because all the
fermion fields $\lt$ are even.

Let us try to construct the partition function by hand.
The partition function for 32 free left-moving fermions is 
\bbb
\hh (I\uu 0\ll 0 + I\uu 1\ll 0)
\eee
in the SO(32) tensor sector
and 
\bbb
\hh (I\uu 0\ll 1 + I\uu 1\ll 1)
\eee
in the SO(32) spinor sector.  Here, $I\uu p\ll q$
is the path integral for fermions on the torus where
the perioditicity is $\lt\to (-1)\uu {q+1} \lt$ on
the spacelike cycle, with a factor of
$(-1)\uu{ p F_{L_W}}$ inserted into the trace,
which translates to a periodicity of $\lt\to (-1)\uu {p+1}\lt$ 
on the timelike cycle.  The numbers $p,q$ are only
meant to be taken mod 2.

\heading{Partition functions for fermions}

In defining our partition functions it is useful
to start by defining partition functions for complex
fermions with various boundary conditions:

\bbb
\bfft \uu 0\ll 0 (\tb) \equiv {1\over{\eta(\tb)}}~
\th\ll{00}(0,\tb)
\\\\
\bfft \uu 1\ll 0(\tb) \equiv {1\over{\eta(\tb)}}~
\th\ll{01}(0,\tb) 
\\\\
 \bfft \uu 0\ll 1(\tb) \equiv {1\over{\eta(\tb)}}~
\th\ll{10}(0,\tb) 
\\\\
\bfft \uu 1\ll 1 (\tb) \equiv {1\over{\eta(\tb)}}~
\th\ll{11}(0,\tb) 
\een{complexfermionpartitionfunctions}
The last of the four vanishes due to
fermion zero modes.  However we will
treat this function as nonzero and demand
modular invariance without using its
vanishing, since the modular invariance of
the partition function must hold
even when the zero modes are lifted
by the insertions of local operators.
These functions have the modular transformation properties
\bbb
\bfft \uu 0\ll 0 (\tb +1)
= \exp{- {{\pi i }\over{12}}}~ \bfft \uu 1\ll 0 (\tb)
\eee \bbb
\bfft \uu 1\ll 0 (\tb +1)
= \exp{- {{\pi i }\over{12}}} ~ \bfft \uu 0\ll 0 (\tb)
\eee \bbb
\bfft \uu 0\ll 1 (\tb +1)
=  \exp{+{{\pi i }\over 4} - {{\pi i}\over{12}}}
~      \bfft \uu 0\ll 1 (\tb)
= \exp{+{{\pi i}\over{6}}}
~      \bfft \uu 0\ll 1 (\tb)
\eee \bbb
\bfft \uu 1\ll 1 (\tb +1)
= \exp{+{{\pi i }\over 4} - {{\pi i}\over{12}}}
~  \bfft \uu 1\ll 1 (\tb)
= \exp{+{{\pi i}\over{6}}}
~      \bfft \uu 1\ll 1 (\tb)
\eee \bbb
\bfft \uu 0\ll 0 (- {1\over\tb})
= ~ \bfft\ll 0\uu 0 (\tb)
\eee \bbb
\bfft \uu 1\ll 0 (- {1\over\tb})
= ~ \bfft\ll 1\uu 0 (\tb)
\eee \bbb
\bfft \uu 0\ll 1 (- {1\over\tb})
= ~ \bfft\ll 0\uu 1 (\tb)
\eee \bbb
\bfft \uu 1\ll 1 (- {1\over\tb})
= -i  \bfft\ll 1\uu 1 (\tb)
\een{complexfermionmodulartransformations}
We have assigned a phase of $-i $ to the
S-transformation of $\bfft \uu 1\ll 1 $ in order
to reproduce the S-transformation of the partition
function for a complex left moving fermion with
insertions.  The phase can be computed with a
single insertion of $\lt\uu 8\lt\uu 9$, a weight $(1,0)$
operator which lifts both fermion zero modes.
Since the $S$-transformation is implemented by
a 90 degree rotatation, it acts with an extra phase of
$i\uu{{\tilde{h} - h}}$ on an operator of weight $(\tilde{h},h)$.
This combines with a $-$ sign from the $\n\to 0$ limit of the
modular transformation of $ {1\over{\eta(\tb)}}~
\th\ll{11}(n,\tb) $ to give an overall phase of
$-i$ in the modular S-transformation of $\bfft\uu 1\ll 1$.

\heading{Fiber partition functions}

Explicit forms for $I\uu p\ll q$ are given by
the theta functions
\bbb
I\uu 0\ll 0 = \qb\uu{- {{16}\over{24}}}
\prod\ll{m = 1}\uu\infty (1 + \qb\uu{m - \hh})\uu{32} = 
\lrdd {{\th\ll{00}(0,\tb)}\over{\eta(\tb)}} \rrdd\uu{16}
\equiv \lrdd \bfft\uu 0\ll 0\rrdd\uu{16}
\\\\
I\uu 1\ll 0  = \qb\uu {- {{16}\over{24}}}
\prod\ll{m = 1}\uu\infty (1 - \qb\uu{m - \hh})\uu{32}
= \lrdd {{\th\ll{01}(0,\tb)}\over{\eta(\tb)}} \rrdd\uu{16}
\equiv \lrdd \bfft\uu 1\ll 0\rrdd\uu{16}
\\\\
I\uu 1\ll 0 = \qb\uu {- {{16}\over{24}} + {{32}\over{16}} }
\prod\ll{m = 0}\uu\infty (1 + \qb\uu{m})\uu{32}
= \lrdd {{\th\ll{10}(0,\tb)}\over{\eta(\tb)}} \rrdd\uu{16}
\equiv \lrdd \bfft\uu 0\ll 1\rrdd\uu{16}
\\\\
I\uu 0\ll 0 = \qb\uu {- {{16}\over{24}} + {{32}\over{16}} }
\prod\ll{m = 0}\uu\infty (1 - \qb\uu{m })\uu{32} = 0
\equiv \lrdd \bfft\uu 1\ll 1\rrdd\uu{16}
\eee
$Z\up R$ is the right-moving partition function, which
is modular invariant up to the local gravitational
anomaly contribution to the $T$ transformation:
\bbb
Z\up R(\t + 1) = \exp{- {{2\pi i }\over 3}}
Z\up R (\t)
\\\\
Z\up R(- {1\over \t}) = Z\up R (\t) 
\eee
The pieces $I\uu p\ll q$
have their 'classical' modular transformations, up
to the contribution to the $T$ transformations:
\bbb
I\uu p\ll q (\t + 1) = 
\exp{+ {{2\pi i }\over 3}}
I\uu{p+q +1}\ll q(\t)
\\\\
I\uu p\ll q(- {1\over \t}) = I\uu{-q}\ll p (\t)
\eee
These can be computed from the modular
transformations of the
$\bfft\uu p\ll q$, given above.

\heading{Base partition functions}

Now we want to correlate the periodicity of states
in the base with their SO(32) spinor number by combining
the $I^p_q$'s with corresponding path integral sectors of
the base. So we let $Y\uu a\ll b$ be the partition function
for the $S\uu 1$ base factor $X\uu 9$ in the sector of the 
path integral where the $X\uu 9$ winds through $2\pi a R\ll{base}$
around the timelike cycle of the worldsheet and $2\pi b R\ll{base}$ 
around the spacelike cycle.  So in particular,
the partition function in the sector with
winding $b$ and momentum $n\ll{base} \equiv p\ll{base} R\ll{base}$ 
equal to $-\th$ mod 1 is given by
\begin{equation}
\label{inverse}
Y\ll b \upb{-\th} \equiv 
\sum\ll a \exp{2\pi i a \th} Y\uu a\ll b =  |\eta(\t)|\uu{-2}
\sum_{n=-\th \bmod 1} {\bar q}^{\frac14p_L^2} q^{\frac14p_R^2}
\end{equation}
where
\begin{equation}
p_L = \frac n{R\ll{base}} -R\ll{base} 
 b\qquad p_R = \frac nR\ll{base}
 + R\ll{base}b
\end{equation}
are left- and right-moving momenta, respectively, and
$q=e^{2\pi i\tau}$, as usual. For instance, $\th = \hh$ would 
correspond to the partition function 
over states with half-integral momentum on the base in units of 
${1\over{R\ll{base}}}$. 

By inverting (\ref{inverse}), one obtains the explicit expression
\begin{equation}
\label{Ys}
Y_b^a = |\eta(\t)|\uu{-2}
\int dn\; e^{2\pi i n a} \bar q^{\frac14p_L^2}
q^{\frac14p_R^2} = |\eta(\t)|\uu{-2}
\frac 1{\sqrt{\tau_2}} e^{(-\pi a^2 -\pi|\tau|^2  b^2 - 
2\pi\tau_1 a b )R\ll{base}\sqd /\tau_2}
\end{equation}
for the $Y_{b}^a$'s. The functions $Y\uu a\ll b$ also have classical
modular transformations:
\bbb
Y\uu a\ll b (\t + 1) = Y\uu{a+b}\ll a (\t)
\\\\
Y\uu a\ll b (- {1\over \t}) = Y\uu {-b}\ll a (\t)
\eee

\heading{Full partition function}

With these ingredients, the partition function in the untwisted (unwound) 
sector of the Wilson line theory can be written as
\bbb
Z = \hh Z\up R~
\sum\ll a Y\uu a\ll 0 (I\uu 0\ll 0 + I\uu 1\ll 0)
+ \hh \sum\ll a (-1)\uu a  Y\uu a\ll 0 (I\uu 0 \ll 1 + I\uu 1\ll 1) 
\eee
For $a$ even, this transforms unproblematically to
itself under the modular $T$ transformation $\t \to \t + 1$
and the modular $S$ transformation $\t\to - {1\over \t}$.
If we write the full partition function in the form $\sum\ll{a,b}
Y\uu a\ll b f\uu a\ll b$, then 
\bbb
f\uu a\ll 0 = \hh Z\up R ~
(I\uu 0\ll 0 + I\uu 1\ll 0 + (-1)\uu a
I\uu 0\ll 1 + (-1)\uu a I\uu 1\ll 1)
\eee
In order for modular invariance to be maintained, the
functions $f\uu a\ll 0$ must also transform classically
under modular transformations.  So
\bbb
f\uu 0\ll b = \hh 
Z\up R ~
(I\uu 0\ll 0 + (-1)\uu b I\uu 1\ll 0 + 
I\uu 0\ll 1 + (-1)\uu b I\uu 1\ll 1)
\eee

By applying modular transformations we find that
\bbb
f\uu 0\ll 1 = \hh Z\up R ~(I\uu 0\ll 0 - I\uu 1\ll 0 + 
I\uu 0\ll 1 -  I\uu 1\ll 1)
\eee
and
\bbb
f\uu 1\ll 1 =  \hh Z\up R ~ (-I\uu 0\ll 0 + I\uu 1\ll 0 + 
I\uu 0\ll 1 -  I\uu 1\ll 1)
\eee
One can check that $f\uu a\ll b$ will depend on $a$ and $b$ 
only mod 2, so we define $f\uu{a + 2k}\ll{b + 2l} = f\uu a\ll b$. 
Given the modular transformations of the $I\uu p\ll q$, the
symbols $f$ then have classical transformation laws
mod $2$ as well.
Fourier transforming back from $a$ to $\th = 0,\hh$, we then find
the partition function in the odd winding sector is
\bbb
Z\up R ~\lrdd Y\uu +\ll b I\uu -\ll 1   +
 Y\uu -\ll b I\uu +\ll 1 \rrdd
\eee
where $Y\uu\pm \ll b\equiv \sum\ll a (\pm)\uu a Y\uu a\ll b$
and $I\uu\pm\ll b$ is defined similarly, with the sum on $a$
running only from $0$ to $1$ and an overall factor of $\hh$ in front.

In terms of states, then we have a theory containing states
with even winding in which the SO(32) tensors have integer momentum
and the SO(32) spinors have half integer momentum, and
states with odd winding in which the SO(32) situation
is the opposite.  Furthermore, the winding mod 2 determines
the $\mflw$ projection.  That is, in states with even winding the projection
is positive (meaning positive-chirality SO(32) spinors and even-rank
SO(32) tensors) and in states with odd winding we have negative-chirality
SO(32) spinors and odd-rank SO(32) tensors.

Note that it is important for level matching that the $\mflw$
projection be reversed in sectors with odd winding 
in the base.  Since the SO(32) tensors have half-integral
momentum, there is a level mismatch in the base of $\hh$ mod 1, and
it must be cancelled by a level mismatch in the fiber
of $\hh$ as well, which comes from the odd number of $\l$-fermions
in these states.

This example illustrates the general concept behind constucting
stringy Wilson lines: we have some path integrals over internal
degrees of freedom $f\uu a\ll b$ with boundary conditions on
the timelike and spacelike cycles labelled by $a$ and $b$, 
and we pair them with the $Y\uu a\ll b$.  Defining the theory
as a Wilson line means choosing the $f\uu a\ll 0$ to implement 
the boundary conditions for bulk fields.  Modular invariance
then determines the rest of the $f\uu a\ll b$ by the requirement
that their modular transformations must be classical, although
the individual summands $I\uu p\ll q$ of $f\uu a\ll b$
need not have classical modular transformations.  (We shall
see that their transformations have anomalous phases in general.)
If we have made a consistent
choice of boundary condition for local fields in the base,
(i.e., a boundary condition
which respects closure of the OPE
on the worldsheet) then there should
be a resulting set of $f\uu a\ll b$ which transforms classically
under the modular $S$ and $T$ transformaions.  This is the
consequence of the CCC principle as discussed in the introduction.

Note, by the way, that some of the internal
path integrals $I\uu p\ll q$ may vanish
(for instance $I\uu 0\ll 0$ in this case),
but we will not use any such vanishings.  The reason is that
when operators are inserted into the path integral, modular invariance
must still hold.  The structure of the partition function with insertions is
exactly the same as the one without insertions, with the new
$I\uu p\ll q$'s
having the modular transformations we have assigned them here, but
no longer vanishing.  

\subsubsection{The case $k\ll L  = 4$}
Next we consider the case $k\ll L  = 4$: 
a chiral discrete symmetry which inverts $4$ of the $\lt$.
The untwisted sector is given by
\bbb
I\uu a\ll 0 = \hh \lrdd (\bfft\uu a\ll 0)\sqd (\bfft\uu 0\ll 0)
\uu{14} + (\bfft\uu {a+1}\ll 0)\sqd (\bfft\uu 1\ll 0)
\uu{14} + (-1)\uu a (\bfft\uu {a}\ll 1)\sqd (\bfft\uu 0\ll 1)
\uu{14}
+ (-1)\uu a (\bfft\uu {a+1}\ll 1)\sqd (\bfft\uu 1\ll 1)\uu{14}
\rrdd ~~~
\eee
These functions are periodic mod 2 in the upper index 
$I\uu{a+2}\ll 0 = I\uu a\ll 0$.  This expresses the
fact that $g$ is a $\IZ\ll 2$ symmetry in the untwisted
sector.  The unprojected partition functions 
in the twisted sectors are
\bbb
I\uu 0\ll b = \hh \lrdd (\bfft\uu 0\ll b)\sqd (\bfft\uu 0\ll 0)
\uu{14} + 
(-1)\uu b
(\bfft\uu {1}\ll b)\sqd (\bfft\uu 1\ll 0)
\uu{14} +  (\bfft\uu 0\ll {b+1})\sqd (\bfft\uu 0\ll 1)
\uu{14}
+  (-1)\uu b
(\bfft\uu {1}\ll {b+1})\sqd (\bfft\uu 1\ll 1)\uu{14}
\rrdd ~~~
\eee
In the twisted NS$_+$ ground state the eigenvalue 
of $g$ is $-1$, so
\bbb
I\uu a \ll b = \hh~(-1)\uu{ab}
~\lrdd  (\bfft\uu a\ll b)\sqd (\bfft\uu 0\ll 0)
\uu{14} + (-1)\uu b ~ (\bfft\uu {a+1}\ll b)\sqd (\bfft\uu 1\ll 0)
\uu{14} \rno
\\\\
\lno + (-1)\uu a ~ (\bfft\uu a\ll {b+1})\sqd (\bfft\uu 0\ll 1)
\uu{14} + (-1)\uu{a+b} ~(\bfft\uu {a+1}
\ll {b+1})\sqd (\bfft\uu 1\ll 1)
\uu{14} \rrdd
\eee
These functions transform under modular transformations
as
\bbb
I\ll b\uu a ( \t + 1) = \exp{{{2\pi i}\over 3}}~\exp{{{\pi i b\sqd}
\over 2}}
~ I\uu{a+b}\ll a (\t)
\\\\
I\ll b\uu a (- {1\over \t}) = (-1)\uu{ab}~I\uu{-b}\ll a (\t)
\eee
In fact $g$ extends to twisted sectors as a $\IZ\ll 2$
symmetry as well, since fermions are odd or even under $g$
in blocks of four, and a state with periodic fermions
gets a phase of $\pm i$ for each two periodic fermions odd
under $g$.  This is reflected in the fact that
$I\uu a\ll b$ is periodic mod 2 in each index separately.

We can construct a unitary, modular invariant partition
function by defining $I\pr{}\uu a\ll b \equiv \exp{{{\pi i a b}
\over 2}} I\uu a\ll b$.  Up to the local
gravitational anomaly, $I\pr{}\uu a\ll b$ transforms
classically under modular transformations:
\bbb
I\pr{}
\uu a\ll b(\t + 1) = \exp{{{2\pi i}\over 3}}I\pr{}\uu {a+b} \ll b
\\\\
I\pr{}\uu a\ll b ( - {1\over \t}) = I\pr{}\uu {-b}\ll a(\t)
\eee
so the combination
\bbb
Z\equiv \sum\ll {ab} Y\uu a\ll b ~I\pr{}\uu a\ll b ~ Z\up R
\\\\
= \sum\ll {ab} Y\uu a\ll b ~I\uu a\ll b ~ Z\up R~ i\uu{ab}
\eee
is modular invariant.  It has an interpretation as a
trace over states:
\bbb
Z =
\sum\ll{a,b}
(q\qb)\uu{-{2\over 3}}~\tr\ll{\rm [twisted~by~ \it g\uu b,~
w\ll{base} = b \rm ] }
\lrdd g\uu a \cdot g\ll X\uu a \cdot
 q\uu {L\ll 0} \qb\uu{\tilde{L}\ll 0} \rrdd~\exp{{{\pi i a b}\over
2}}
\eee
which amounts to a projection onto states with
$g\cdot g\ll X = \exp{- {{\pi i w\ll{base}}\over 2}}$.  Since
as we argued above $g = \pm 1$ for all states, twisted
and untwisted, it follows that $g\ll X \equiv \exp{2\pi i p\ll
{base} R\ll {base}}$ must have eigenvalues $\pm i$ when $w\ll{base}$
is odd.  In other words,
$p\ll{base}$ is fractionated in twisted sectors
in units of ${1\over{4 R\ll{base}}}$.  This matches the
value one would derive by demanding level matching in
the sector $w\ll{base} = 1$.

\subsubsection{Other values of $k\ll L \equiv 4 m$}

In general\footnote{Our $k\ll L = 0$ case in
an earlier sub-subsection does not fall into
this set -- in that case we had an extra action of $g$
on $SO(32)$ spinor number in order to get a nontrivial
action for $g$ at all.}, we have
\bbb
I\uu a \ll b \equiv \hh~(-1)\uu{mab}~\lrdd
(\bfft\uu a\ll b)\uu {2m} (\bfft\uu 0\ll 0)\uu{16-2m}
+ (-1)\uu{mb} (\bfft\uu {a+1}
\ll b)\uu {2m} (\bfft\uu  1\ll 0)\uu{16-2m} 
\rno
\\\\
\lno 
+ (-1)\uu{ma} (\bfft\uu {a}\ll {b+1})\uu{2m}
(\bfft\uu  0\ll 1)\uu{16-2m} 
+ (-1)\uu{m(a+b)}
(\bfft\uu {a+1}
\ll {b+1})\uu {2m}
 (\bfft\uu 1\ll 1 )\uu{16-2m}
\rrdd
\eee

\bbb
I\ll b\uu a ( \t + 1) = \exp{{{2\pi i}\over 3}}~\exp{{{\pi im b\sqd}
\over 2}}
~ I\uu{a+b}\ll a (\t)
\\\\
I\ll b\uu a (- {1\over \t}) = (-1)\uu{mab}~I\uu{-b}\ll a (\t)
\eee
In fact $g$ extends to twisted sectors as a $\IZ\ll 2$
symmetry as well, since fermions are odd or even under $g$
in blocks of four, and a state with periodic fermions
gets a phase of $\pm i$ for each two periodic fermions odd
under $g$.  This is reflected in the fact that
$I\uu a\ll b$ is periodic mod 2 in each index separately.

We can construct a unitary, modular invariant partition
function by defining $I\pr{}\uu a\ll b \equiv \exp{{{\pi im a b}
\over 2}} I\uu a\ll b$.  Up to the local
gravitational anomaly, $I\pr{}\uu a\ll b$ transforms
classically under modular transformations:
\bbb
I\pr{}
\uu a\ll b(\t + 1) = \exp{{{2\pi i}\over 3}}I\pr{}\uu {a+b} \ll b
\\\\
I\pr{}\uu a\ll b ( - {1\over \t}) = I\pr{}\uu {-b}\ll a(\t)
\eee
so the modular invariant combination is
\bbb
Z\equiv \sum\ll {ab} Y\uu a\ll b ~I\pr{}\uu a\ll b ~ Z\up R
= \sum\ll {ab} Y\uu a\ll b ~I\uu a\ll b ~ Z\up R~ i\uu{mab}.
\eee
The partition function $Z$ has an interpretation as a
trace over states:
\bbb
Z =
\sum\ll{a,b}
(q\qb)\uu{-{2\over 3}}~\tr\ll{\rm [twisted~by~ \it g\uu b,~
w\ll{base} = b \rm ] }
\lrdd g\uu a \cdot g\ll X\uu a \cdot
 q\uu {L\ll 0} \qb\uu{\tilde{L}\ll 0} \rrdd~\exp{{{\pi i m a b}\over
2}}
\eee
which amounts to a projection onto states with
$g\cdot g\ll X = \exp{- {{\pi i m w\ll{base}}\over 2}}$.  Since
as we argued above $g = \pm 1$ for all states, twisted
and untwisted, it follows that $g\ll X \equiv \exp{2\pi i p\ll
{base} R\ll {base}}$ must have eigenvalues 
$\pm i\uu m$ when $w\ll{base}$
is odd.  $p\ll{base}$ is fractionated
in units of ${m\over{4 R\ll{base}}}$ mod
${1\over {R\ll{base}}}$.  The
Wilson line is tame when $m$ is even and wild when
$m$ is odd.

\subsection{Wilson line for T-duality in the bosonic string}

Now we turn to one of the more interesting and 'stringy'
symmetries of the bosonic string compactified on
a fiber circle of self-dual radius $\sqrt{\apr}$.  As is
well known, this theory has a duality symmetry which
maps the theory to itself with an exchange of
modes carrying momentum $n$ and winding $w$ on the fiber
circle.  We would like to compactify a second dimension
on a circle to play
the role of a base, and impose a Wilson line boundary condition
such that all string states undergo a T-duality transformation
when transported around the base.  

Such a
theory would be an example of a T-fold (\cite{hull2}).
Much
work has been done on the construction of
T-folds at the level of effective field theory in the
base and also
on some aspects of the worldsheet realizations
(\cite{HMW,STW, FWW, LSW, hull1, hull2, 
gray, hull3, hull4, hull5}).
However an explicitly modular invariant
worldsheet description is lacking for
even some of the simplest T-folds \cite{FW}.
In this section and subsequent sections
we will demonstrate how the worldsheet CFT construction of 
a string propagating on a T-fold or more
general monodrofold (\cite{FWW}) can be carried out in general.

In all that
follows, ${\cal I}$ will be the theory of
a circle $S\uu 1$ compactified at the
self-dual radius in the bosonic string. In this subsection, we
give a direct discussion of the Wilson line theory, along the
lines of the previous example. In the next subsection, we will
identify several alternative derivations of the same result.

\heading{Action of T-duality symmetry in the fiber ${\cal I}$}

In order to make this construction, it is important
to make one observation about the action of
T-duality which appears to have gone unremarked in
the literature.  The technical
basis for this observation is explained in detail
in the Appendix.

The naive action 
\bbb
T\ll{naive} \kket{n,w} = \kket{w,n}
\eee
which simply reverses the momentum and winding of a state
is not actually a good symmetry of the two-dimensional
CFT, in the sense that the operator product expansion
does not preserve it.  For instance, it is possible
to take the OPE of two states which have eigenvalue
$+1$ under $T\ll{naive}$ and get a state on the
right hand side of the OPE which has eigenvalue $-1$.
The reason for this has to do with the zero-mode cocycle
factors which appear in the definition of
vertex operators carrying momentum and winding in
a toroidally compactified theory.  There is still
a true T-duality symmetry which switches momentum $n$ and
winding $w$, with an extra phase depending on
$n$ and $w$:
\begin{equation}
\label{TTrue}
T\ll{true} \kket{n,w} = (-1)\uu{nw} \kket{w,n}
\end{equation}
For any state with $n\neq w$ this phase can
be absorbed into a redefinition of 
the phase of the state.  However for states with
$n = w$ the extra phase is meaningful.  The
naive T-duality operation always has positive eigenvalues
for states with $n = w$, whereas the true T-duality operation
has phase $+1$ when $n=w$ is even and $-1$ when $n=w$ is odd.

{} From here on, when we refer to T-duality we will always 
mean the true, conserved T-duality operation which
switches $n$ with $w$ and acts with a minus sign on
all left-moving oscillators $\tilde{\a}\ll n$
in the self-dual $c=1$ circle CFT.

\heading{Twisted sector}

Now we would like to make three more observations, relating
to the properties of the twisted sectors.
In a twisted sector, the left-moving current $\pp\ll - X$
is antiperiodic, and therefore
all left-moving oscillators $\tilde{\a}\ll{n + \hh}$
are half-integrally moded.

In the twisted sector
the left moving modes have a Casimir energy equal to ${1\over{16}}$,
and it is not even clear that we \it should \rm expect
the twisted sector to be level-matched.  That is,
there is no particular reason to expect an
asymmetric orbifold by $\hat{T}$ to be consistent, so
there is no \it a priori \rm necessity of modular invariance
which would demand level matching in the twisted sector.
However we can forge ahead and explore the hypothesis that
the orbifold by $\hat{T}$ may be modular invariant, seeing
where this leads us.

The only possible source of fractional
energy on the right would be a nontrivial
quantization for the zero mode in the twisted sector.
This is a natural possibility to consider, as the
zero mode quantization condition $p\ll L - p\ll R \in 2\IZ$
has no analog in the untwisted sector.

In order for the
sector twisted by $\hat{T}$ to be level matched,
the twisted ground state would have
to have $p\ll R$ equal to $\pm\hh$ (mod 4), in order that
the right-moving energy ${1\over 4}p\ll R\sqd$
should equal ${1\over{16}}$.  We will assume that
this is the correct zero mode quantization in the twisted
sectors.





Now to our second comment about
the twisted sectors.
In order for T-duality to
act in a consistent way in the twisted sector, it must
commute with T-even operators in ${\cal I}$
and anticommute with T-odd operators, in their action on
twisted states.  For this to work out, there must be
a nontrivial cocycle factor in
the action of $T$ on the twisted sector which
depends on the zero mode $p\ll R$, analogous to the cocycle
phase $(-1)\uu{nw}$ in the untwisted states ${\cal I}$.
In the Appendix we work out a cocycle contribution
to $T$ in the twisted sectors with the requisite properties.

Our third and last point
is that for twisted states there is no mod 2 condition
\it per se \rm on the value of $p\ll R$.  That is, if there
exists a twisted state ${\cal T}$
with some value of $p\ll R$, it is 
always possible to find an untwisted operator ${\cal O}
\in {\cal I}$ with $p\ll R\pr = p\ll R + 1$, and the product 
${\cal T}\cdot {\cal O}$ is a twisted state with the same
eigenvalue of $T$ and the opposite value of $\exp{\pi i p\ll R}$.

Though it may seem surprising that
there are 'fewer' constraints on $p\ll R$ mod 2
in the twisted sector than in the untwisted sector,
there is a point of view from which it is less mysterious.
The zero mode of $X\uu L$
is nondynamical, but it may nonetheless act as a discrete label which
carries no energy and labels
distinguishable but degenerate twisted sectors.
These
labels are quite familiar from geometric orbifold compactifications,
where they simply label the fixed points of a
geometric symmetry.  In general they can
be thought of as distinct eigenspaces for the 
zero modes of the orbifolded fields.  In the twisted sectors,
then, we can if we like consider states with $p\ll R = +\hh$
and those with $p\ll R = - \hh$ to be paired with
distinct left-moving twisted sectors, distinguished by
nondynamical labels of the left-moving zero modes.
Then we can by fiat extend the quantum number $\exp{\pi i p\ll L}$
as a quantum number acting on twisted states, such
that $\exp{\pi i p\ll L} = \exp{\pi i p\ll R}$ for \it all \rm
sectors, twisted and untwisted.

In the Appendix we construct the fiber
partition functions of
the asymmetric orbifold by $T$ in the
bosonic string.  We find that
the correct $p\ll R$ projection in the twisted sectors
is the vacuous one: for a $T$-even state of the
left-moving degrees of freedom, both values $\pm \hh$
of $p\ll R$ are allowed.  Likewise, for a $T$-odd state
on the left,
both values $\pm {3\over 2}$ are allowed.


The partition functions of the
fiber theory are derived in the Appendix.  In the
T-even untwisted sector we have
\bbb
I\uu +\ll 0  \equiv \hh \lrdd I\uu 0\ll 0 + I\uu 1\ll 0 \rrdd
\eee
and in the T-odd untwisted sector we have
\bbb
I\uu -\ll 0  \equiv \hh \lrdd I\uu 0\ll 0 - I\uu 1\ll 0 \rrdd
\eee
with
\bbb
I\uu 0\ll 0 = 
|\eta(\t)|\uu{-2} ~\lrdd |\th\ll{00} (0,2\t)|\sqd
+ |\th\ll{10} (0,2\t)|\sqd \rrdd 
\\\\
I\uu 1\ll 0 = \eta(\t)\uu{-1} ~\lrdd {{2\eta(\tb)}\over
{\th\ll{10}(0,\tb)}}\rrdd\uu\hh ~ 
\th\ll{01} (0,2\tb) 
\eee
where $I\uu a\ll 0$ represents the partition function in
the untwisted sector with an insertion of $T\uu a$. 
That is, $I\uu 0\ll 0$ is simply the partition function for
a $S\uu 1$ CFT at the self-dual radius.

Similarly, in the twisted sector the
partition functions are
\bbb
I\uu 0\ll 1 =  \eta(\t)\uu{-1} ~\lrdd {{\eta(\tb)}\over
{\th\ll{01}(0,\tb)}}\rrdd\uu\hh ~ 
\th\ll{10} (0,{{\tb\over 2}}) 
\\\\
I\uu 1\ll 1 = \sqrt{2}~\eta(\t)\uu{-1} ~\lrdd {{\eta(\tb)}\over
{\th\ll{00}(0,\tb)}}\rrdd\uu\hh ~ 
\th\ll{10}( - {1\over 4}, {{\tb}\over 2})
\eee
and
\bbb
I\uu \pm \ll 1 = \hh \lrdd I\uu 0\ll 1 \pm I\uu 1\ll 1 \rrdd
\eee

The modular transformations of these objects are
simply the classical ones, $I\uu a\ll b (\t + 1)
= I\uu{a+b}\ll b(\t)$ and $I\uu a\ll b (- {1\over \t})
= I\uu {-b} \ll a$, where $I\uu a\ll b$ is
defined to be periodic mod 2 in $a$ and $b$ separately.

This means in particular that
it is simple to define a partition function for the
asymmetric \it orbifold \rm (as opposed to
Wilson line) $S\uu 1 / T$:
\bbb
Z\uu{asym.~orb.} \equiv
I\uu +\ll 0 + I\uu +\ll 1
\\\\
= \hh \lrdd I\uu 0\ll 0 + I\uu 1 \ll 0 + I\uu 0\ll 1 + I\uu 1 \ll 1
\rrdd
\eee
which is unitary and modular invariant.


In order to define the Wilson line for $T$ over a circle of
radius $R$, define $Y\uu a\ll b$ to be the partition
function in which the base coordinate $X\ll{base}$ winds
$a$ times around the Euclidean timelike direction
of the worldsheet, and $b$ times around the
spacelike direction of the worldsheet.  This
function is related to the unitary 
partition functions (\ref{inverse}) for $S\uu 1$ by
\bbb
Y\uu a\ll b \equiv  \int\ll 0\uu 1 d\th
\exp{2\pi i a \th} Y\ll b\upb{-\th},
\eee
where $Y\ll b \upb {\th}$ is the partition function for the base
in the sector with winding $w\ll{base}$ equal to $b$, and
momentum $n\ll{base} \equiv R\ll{base} p\ll{base}$ equal
to ${{\th}\over{2\pi}}$ mod 1.
The functions $Y\uu a\ll b$ also have classical
modular properties
\bbb
Y\uu a\ll b (\t + 1) = Y\uu{a+b}\ll b(\t)
\eee
and
\bbb
Y\uu a\ll b(- {1\over\t}) = 
Y\uu {-b}\ll a(\t)
\eee
The partition function for the Wilson line is
\bbb
Z\uu{Wilson~line} =
\sum\ll{b}
Y\ll b\uu + I\ll b\uu + + Y\ll b\uu -  I\ll b\uu -
\eee
where as above $I\ll b\uu \pm \equiv \hh \lrdd 
I\ll b\uu 0 \pm I\ll b\uu 1 \rrdd $
and
\bbb
Y\ll b\uu \pm \equiv \sum \ll a (\pm 1)\uu a Y\ll b\uu a
\\\\
= Y\ll b\upb{{1 \over 4}(1 \pm 1)} 
\eee
With these definitions, we can rewrite our
full partition function as
\bbb
Z\uu{Wilson~line} =
\hh\sum\ll{a,b} Y\ll b\uu a I\ll b\uu a
\eee
Since both $Y\uu a\ll b$ and $I\uu a\ll b$
transform classically under modular transformations,
the partition function $Z\uu{Wilson~line}$
is modular invariant.

In the sector of zero
winding $b = 0$ the sum is
\bbb
Z\ll{w = 0}\equiv
\sum\ll a Y\uu a\ll 0 I\uu a\ll 0
\\\\
=  \lrdd \hh \sum\ll a  Y\uu a \ll 0 \rrdd ~I\uu +\ll 0
+ \lrdd \hh \sum\ll a (-1)\uu a Y\uu a\ll 0 \rrdd ~I\uu -\ll 0
\eee
which is precisely the sum over states in
which the eigenvalue of $T$ is equal to $\exp{\pi i 
p\ll{base} R\ll{base}}$.  The partition function
is explicitly unitary, modular invariant and 
satisfies our definition of a Wilson line CFT.  The
Wilson line is self-evidently tame, a result of the
classical, anomaly-free modular transformations
of the internal partition functions $I\uu a\ll b$.

Having constructed the Wilson line CFT for T-duality
on a single $S\uu 1$ of self-dual radius, it is
straigtforward to construct Wilson line theories
in which parallel transport around the base
induces a simultaneous T-duality on $k$ self-dual
$S\uu 1$ factors at once.  The internal
partition functions are simply raised to the $k\uu{\rm
{\underline{th}}}$ power:
\bbb
Z \equiv \sum\ll{a,b}  Y\uu a\ll b \lrdd I\uu a\ll b
\rrdd \uu k 
\een{morefibers}
The Wilson line is unitary, modular invariant, and tame for
all values of $k$.

Later we shall study a version of this Wilson line
theory (with $k=2$) in the type II
superstring.  We shall see that the type II
Wilson line theory is still perfectly consistent, but
no longer tame.



\subsection{More on T-duality Wilson line and equivalence with
other models}

\def\zet{\mathbb Z}

We will now reinterpret the discussion from the previous subsection
in a somewhat more abstract language, starting from the chiral algebra
of the circle at the self-dual radius $R_{\rm self-dual}=1$ and its 
$\zet_2$ orbifold \cite{DVVV}.

\heading{Asymmetric orbifold by T-duality}

In general, an orbifold model is defined by specifying the action of the 
symmetry group on the Hilbert space of the parent theory. For the free
boson at the self-dual radius, this Hilbert space is obtained by the
action of the $SU(2)_L\times SU(2)_R$ chiral algebra generated by
\begin{equation}
J_3 = \del X_R\,, \qquad J^\pm = e^{\pm 2 i X_R}\,; \qquad\qquad
\bar J_3 = \delbar X_L\,, \qquad \bar J^\pm = e^{\pm 2 i X_L}
\end{equation}
on the vacuum $|0\rangle$ and the non-trivial primary 
$e^{i(X_R+X_L)}|0\rangle$. In this language, the naive action of
T-duality is as inversion of one of the two $U(1)$ current
\begin{equation}
T: \del X_R \to \del X_R \qquad
\delbar X_L \to - \delbar X_L 
\eqlabel{Tduality}
\end{equation}
But as we have seen in the previous subsection, this is not the
entire story. It is not hard to see that the ``correct'' action
(\ref{TTrue}) which is required to preserve the OPE, must act on 
the full chiral algebra as
\begin{equation}
T:\quad J_3\to J_3\,,\qquad J^\pm \to -J^{\pm}\,; \qquad\qquad
\bar J_3\to -\bar J_3\,, \qquad \bar J^\pm\to - \bar J^{\mp}
\end{equation}
This is the actual action of T-duality on the full chiral algebra
of the free boson at the self-dual radius.
Introducing $J_2=\cos 2X_R$ and $J_1=\sin 2 X_R$, 
$\bar J_2=\cos 2X_L$, $\bar J_1=\sin 2X_L$, $T$ acts as
\begin{equation}
J_3\to J_3\,,\quad J_2\to -J_2\,,\quad J_1\to -J_1\,; \quad\qquad
\bar J_3 \to -\bar J_3\,, \quad \bar J_2\to -\bar J_2\,, \quad \bar J_1\to\bar J_1
\eqlabel{trueT}
\end{equation}
It is then not hard to see that \eqref{trueT} is equivalent by a chiral
$SU(2)$ rotation on the right to the standard, right-left symmetric reflection of 
the circle $X\to -X$. As a consequence, \it the asymmetric orbifold by
T-duality is equivalent, as an abstract conformal field theory, to the
standard symmetric orbifold of the boson at the self-dual radius. \rm
Nevertheless, T-duality and the standard reflection are distinguished after
breaking the $SU(2)_R\times SU(2)_L$ symmetry by deforming away from the 
self-dual radius.

Given this (non-trivial) equivalence between the asymmetric orbifold and the
standard orbifold, it is straightforward to explicitly compute the 
partition function and demonstrate modular invariance. One recovers the
results from the previous subsection. We also find it instructive to 
describe the twisted sectors in the asymmetric 
language. While on the left, we have the standard twist fields (we will 
mostly follow the canonical conventions of \cite{DVVV}), one finds that
the right-movers have a {\it fractional momentum} with respect to $J_3$,
which is quantized in units of $1/2R_{\rm self-dual}$ instead of 
$1/R_{\rm self-dual}$, with the quantum depending on the twisted sector.

In hindsight, this is not surprising. Recall that the standard orbifold
of $S^1_{R_{\rm self-dual}}$ is also equivalent to the standard circle
at twice the self-dual radius $R_{\rm orbifold}=2R_{\rm self-dual}=2$, 
with an equivalence of chiral sectors given in table \ref{equivalence}.
The novelty is that besides viewing this theory as the boson at 
$R_{\rm orbifold}=2$, or as the orbifold of the boson at the self-dual 
radius, one can also view it as the asymmetric orbifold by T-duality. In 
this point of view, the left-movers live on $S^1_{R_{\rm self-dual}}/\zet_2$,
and the right-movers live on $S^1_{R_{\rm self-dual}}$, albeit with fractional 
momentum in the twisted sectors.\footnote{In this section, we use RCFT 
conventions. For a circle at radius $R$ we have chiral momenta $k_{L,R}=
R p_{L,R}$, which are related to integrally quantized momentum and winding by
$$
k_{R,L} = n \pm R^2 w 
$$
and the conformal weights are $\Delta_{L,R}= k_{L,R}^2/ 4R^2$. When
$R^2=r/s$ is rational, the chiral algebra is extended by fields with
$(n,w)=(r,s)$ of dimension $rs$, and there are $2 r s$ primary fields 
labelled by $s k_{L,R}\in\zet\bmod 2 rs$.}

\begin{table}[h]
\begin{center}
\begin{tabular}{|c|c|}
\hline
(chiral) boson sector & (chiral) orbifold sector \\
labelled by momentum $k\bmod 8$ &  \\\hline
$0$ & ${\rm id}$ \\
$4$ & $J$ \\
$2$ & $\bigl(\frac 14\bigr)_+$ \\
$-2$ & $\bigl(\frac 14\bigr)_-$ \\
$\pm 1$ & $\sigma_{1,2}$ \\
$\pm 3$ & $\tau_{1,2}$ \\
\hline
\end{tabular}
\caption{Equivalence of chiral sectors between orbifold 
$S^1_{R_{\rm self-dual}}/\zet_2$ and circle theory 
$S^1_{R_{\rm orbifold}}$. Notations follow \cite{DVVV}.}
\label{equivalence}
\end{center}
\end{table}

\heading{Wilson line CFT}

We can now define the fibration of $S^1_{R_{\rm self-dual}}$ over 
$S^1_L$ with a Wilson line for T-duality as the asymmetric orbifold 
of  $S^1_{\rm R_{\rm self-dual}}\times S^1_{2R}/\zet_2$, where the
$\zet_2$ acts on the first factor as in \eqref{trueT} and on the base 
by a half-shift around the circle. We obtain table \ref{sectors},
in which the base sectors are labelled by momentum and winding, and the
fiber sectors are labelled on the left by the chiral momentum and
on the left by the orbifold sector (the latter two sets of labels 
are of course equivalent by table \ref{equivalence}).

\begin{table}[th]
\begin{center}
\begin{tabular}{|c|c|c|}
\hline
sector & base momentum, winding & fiber  \\
& $(n^b,w^b)\bmod (1,2)$ &  $(k_R^f\bmod 8,{\rm orbifold\; sector})$ \\\hline
untwisted even & $(0,0)$ & $(0,{\rm id}), (2,(1/4)_+), (4,J), (-2,(1/4)_-)$ \\
untwisted odd & $(\frac12,0)$ & $(0,J), (2,(1/4)_-), (4,0), (-2,(1/4)_+)$ \\
twisted even & $(0,1)$ & $(1,\sigma_1), (3,\tau_1), (-3,\tau_2), (-1,\sigma_2)$ \\ 
twisted odd & $(\frac12,1)$ & $(1,\tau_2), (3,\sigma_2), (-3,\sigma_1), (-1,\tau_1)$
\\\hline
\end{tabular} 
\caption{Sectors of the CFT description of the Wilson line for T-duality.}
\label{sectors}
\end{center}
\end{table}

It should be clear that since our orbifold action preserves a $U(1)^2$ 
chiral symmetry on both the right and on the left, this theory is equivalent 
to a particular point in the moduli space of a two-dimensional torus. The 
brute force method to determine this point is to identify the Narain lattice 
of charges with respect to the $U(1)_R^2\times U(1)^2_L$ current algebra. 
In notation explained in the last footnote, we find
\begin{multline}
(k_R^f,k_R^b,k_L^f,k_L^b) \in 
(8,0,0,0) \zet + (0,1,0,1) \zet + (0,2 R^2,0,-2R^2) \zet +
(2,0,2,0) \zet \\ + (4,1/2,0,1/2) \zet + (1,R^2,1,-R^2) \zet \\
\cong (8,0,0,0)\zet + (2,0,2,0)\zet+(4,1/2,0,1/2)\zet+
(1,R^2,1,-R^2)\zet
\end{multline}
and the right/left metric
\begin{equation}
4\Delta_{L,R} = \frac{\bigl(k_{L,R}^f\bigr)^2}4 + 
\frac{\bigl(k_{L,R}^b\bigr)^2}{R^2}
\end{equation}
The standard parametrization of the momentum lattice at a point in the moduli 
space of the torus is
\begin{equation}
\begin{split}
q_{L,R} &= n - B w \pm g w \\
4 \Delta_{L,R} &= g^{-1} (q_{L,R},q_{L,R}) 
\end{split}
\end{equation}
Here $g$ is the metric and $B$ the B-field on the torus, and $n$
and $w$ are now two-dimensional momentum and winding, respectively.

After a little bit of algebra, we find that the two lattices are
isomorphic precisely if the moduli of the torus take the values
(up to an $SL(2,\zet)\times SL(2,\zet)$ transformation)
\begin{equation}
\rho = \frac{i}R \qquad \tau = \frac12+\frac i{4 R}
\label{result}
\end{equation}

There is, of course, a more economical way to derive this result. 
Namely, we notice that T-duality is not only
equivalent to the symmetric reflection, but also to half a
shift around the circle at the self-dual radius. Therefore,
our Wilson line CFT is an orbifold of two circles, one of radius 
$2 R_{\rm base} = 2R$, and one of radius $R_{\rm fiber}=1$, by the 
action
\begin{equation}
(X^f,X^b) \to (X^f+1/2,X^b+1/2)
\end{equation}
It is easy to see that the resulting torus has moduli given
by \eqref{result}.

Finally, we note that we can also view the Wilson CFT as the orbifold 
of $S^1_{R_{\rm self-dual}}\times S^1_{2R}$ by
\begin{equation}
(X^f,X^b)\to (-X^f,X^b+1/2)
\end{equation}
which we recognize as the Klein bottle (as target space). This implies the 
amusing fact that the torus CFT at the point \eqref{result} in moduli space 
is equivalent to a Klein bottle CFT. The moduli space of the Klein bottle
CFT is two-dimensional and it is not hard to see that the intersection
locus of torus and Klein bottle moduli space is just the one-dimension
here parametrized by $R$.

\subsection{Wilson lines in an $SU(2)$
current algebra at level $k$}

Going beyond free field theories, it is natural to test the set of ideas 
presented in this paper in the situation in which the fiber theory is 
described by a rational conformal field theory, such as a WZW model or
a coset model. Such theories have a tendency to possess a host of discrete 
symmetries which can and have been used in a variety of different 
contexts. It is consequential to explore wilsonization of such
symmetries.

We will here examine the simplest such example in detail, when the fiber 
theory is the $SU(2)$ WZW model at level $k$, and the symmetry which we
want to wilsonize originates from chirally acting charge conjugation.
The requisite representation theoretic data is most easily extracted from
ref.\ \cite{bifs}, which treats $\zet_2$ orbifolds of WZW models in general
and which will hence also be a useful input in constructing Wilson line 
theories for such symmetries.

\def\cH{{\mathcal H}}
\def\zbar{\bar z}
The chiral algebra of the model of our interest is the $SU(2)$ current algebra 
generated by $J_3(z), J^\pm(z)$ on the left, and $\bar J_3(\zbar), \bar 
J^\pm(\zbar)$ on the right. The finite number of primary fields are in 
correspondence with the finite number of integrable representations, 
$\cH_l$ of the $\widehat{su}(2)_k$ affine Lie algebra and are labelled 
by twice the spin $l=0,1,\ldots,k$. The modular invariant torus partition 
function is built from the chiral blocks
$\chi_l(\tau)=\tr_{\cH_l} q^{L_0-c/24}$ and given by
\def\chibar{\bar\chi}
\def\taubar{\bar\tau}
\begin{equation}
\boi_0^0(\tau) = \sum_{l=0}^k \chibar_l(\taubar)\chi_l(\tau)
\end{equation}
The symmetry $g$ we wish to wilsonize originates from the $\zet_2$ automorphism
of the chiral algebra which acts on the generators in the chiral fashion
\footnote{In this section we act with our discrete chiral symmetry
on the 
right rather than on the left.}
\begin{equation}
\label{auto}
g: \qquad J_3 \to -J_3, \qquad J^\pm \to J^\mp, \qquad
\bar J_3\to\bar J_3, \qquad \bar J^\pm \to \bar J^\pm
\end{equation}
and can be implemented on the highest weight modules as explained
in \cite{bifs}. For $SU(2)$, charge conjugation is an inner automorphism,
and (\ref{auto}) is in fact nothing but a chiral rotation by $\pi$ around 
the $y$-axis. Since spinor representations pick up a minus sign under a $2\pi$
rotation, as in many examples in the previous subsections, the true
symmetry of the OPE is a $\zet_4$ extension\footnote{which we will abusively 
continue to denote by $g$. \rm}
of $\zet_2$ charge conjugation by the 
$\zet_2$ center of $SU(2)$. In other words, we are implementing
$g$ such that $g^2=(-1)^l$ on $\cH_l$.

The twisted partition functions $\boi_a^b$ of the fiber theory can be
expressed in terms of twisted chiral blocks. We introduce
\begin{equation}
\chi_l^g (\tau) = \tr_{\cH_l} g q^{L_0-c/24}
\label{twin}
\end{equation}
for the trace of $g$ in the untwisted sector as well as
\begin{equation}
\tilde\chi_l(\tau) = \tr_{\cH_l} q^{\tilde L_0-c/24}
\label{twist}
\end{equation}
for the twisted representations. Finally,
\begin{equation}
\tilde\chi_l^g(\tau) = \tr_{\cH_l}g q^{\tilde L_0-c/24}
\label{twisttwin}
\end{equation}
Explicit expressions for these characters are as follows \cite{bifs}
\begin{equation}
\label{explicit}
\begin{array}{r@{\;}c@{\;}l@{\qquad\quad}r@{\;}c@{\;}l}
\chi_l(\tau) &=& \chi_l[0,0](\tau,0) & 
\chi_l^g(\tau) &=& \chi_l[0,1/2](\tau,0) \\
\tilde\chi_l(\tau) &=& \chi_l[1/2,0](\tau,0) &
\tilde \chi_l^g(\tau) &=& e^{-2\pi i k/8} \chi_l[1/2,1/2](\tau,0)
\end{array}
\end{equation}
where
\begin{equation}
\chi_l[s_1,s_2](\tau,z) = 
\frac{\Theta_{l+1,k+2}[s_1,s_2](\tau,z)-\Theta_{-l-1,k+2}[s_1,s_2](\tau,z)}
{\Theta_{1,2}[s_1,s_2](\tau,z)-\Theta_{-1,2}[s_1,s_2](\tau,z)}
\end{equation}
and
\begin{equation}
\Theta_{j,h}[s_1,s_2](\tau,z) = 
\sum_{n=-\infty}^\infty
e^{2\pi i \tau(j + hs_1 + 2 n h)^2/4h} e^{2\pi i(z+s_2)(j + hs_1+ 2 n h)/2}
\end{equation}
is the twisted theta function associated with the root lattice of 
$\widehat{su}(2)$.

The modular transformation properties of the twisted characters are
straightforward to work out. Some of the more important ones are
\begin{equation}
\begin{array}{r@{\;}c@{\;}l@{\qquad\quad}r@{\;}c@{\;}l}
\chi_l(\tau+1) &=& T_l\chi_l(\tau) &
\chi_l^g(\tau+1) &=& T_l\chi_l^g(\tau) \\
\tilde\chi_l(\tau+1) &=& \zeta T_l\tilde\chi_l^g(\tau) & 
\tilde\chi_l(\tau+2) &=& (-1)^l \zeta^2 T_l^2 \tilde\chi_l(\tau)\\
\chi_l(-1/\tau) &=&  S_{ll'} \chi_{l'}(\tau) &
\chi^g_l(-1/\tau) &=&  S_{ll'} \tilde\chi_{l'}(\tau) \\
\tilde\chi_l(-1/\tau) &=&  S_{ll'} \chi_{l'}^g(\tau) &
\tilde\chi_l^g(-1/\tau) &=& (-1)^l\zeta^2 S_{ll'}\tilde\chi_{l'}^g(\tau)
\end{array}
\end{equation}
where $T_l=e^{2\pi i [ l(l+2)/4(k+2) - k/8(k+2)]}$ and
$S_{ll'}=\sqrt{2/h} \sin\bigl[\pi (l+1)(l'+2)/(k+2)\bigr]$ are,
respectively, modular T- and S-matrices of $\widehat{su}(2)$ 
at level $k$, and a sum over $l'$ is understood in all
S-transformations. The phase $\zeta=e^{2\pi i k/16}$ is for general
$k$ a $16$-th root of unity and is the fundamental obstruction to
the existence of the orbifold by $g$.

In terms of these chiral blocks, the explicit expressions for the torus 
partition functions with $g^a$ twist in the space direction and 
$g^b$ twist in the time direction are\footnote{Again, a sum over
$l$ is implied and suppressed.}
\begin{equation}
\label{blocks}
\begin{array}{l<{\quad}l<{\quad}l<{\quad}l}
\boi_0^0 = \chibar_l\chi_l &
\boi_0^1 =  \chibar_l\chi^g_l &
\boi_0^2 =  \chibar_l(-1)^l\chi_l &
\boi_0^3 =  \chibar_l(-1)^l\chi^g_l  \\ 
\boi_1^0 = \chibar_l\tilde\chi_l &
\boi_1^1 =\chibar_l\tilde\chi_l^g &
\boi_1^2 = \chibar_l(-1)^l\tilde\chi_l  &
\boi_1^3 = \chibar_l(-1)^l\tilde\chi_l^g \\
\boi_2^0 = \chibar_l\chi_{k-l} &
\boi_2^1 = \chibar_l\chi_{k-l}^g &
\boi_2^2 = \chibar_l(-1)^{k-l}\chi_{k-l} &
\boi_2^3 = \chibar_l(-1)^{k-l}\chi_{k-l}^g\\
\boi_3^0 =\chibar_l\tilde\chi_{k-l} &
\boi_3^1 = \chibar_l\tilde\chi_{k-l}^g &
\boi_3^2 = \chibar_l(-1)^{k-l}\tilde\chi_{k-l} &
\boi_3^3 = \chibar_l(-1)^{k-l}\tilde\chi_{k-l}^g
\end{array}
\end{equation}
It is straightforward to work out the modular transformation
properties of these expressions. As expected, we cannot in general
form an invariant expression by combining only the $\boi_a^b$'s. The
fundamental obstruction is the level mismatch in the first twisted
sector
\begin{equation}
\boi_1^0(\tau+4) = \zeta^4 \boi_1^0(\tau) \qquad\qquad (\zeta=e^{2\pi i k/16})
\end{equation}
The reason that for $\zeta^4\neq 1$, we cannot project onto $g$ 
invariant states by summing over the T-orbit of $\boi_1^0(\tau)$ is that 
in these cases $g$ does not extend to a true symmetry of the 
OPE in the twisted sectors.

Let us clarify this statement. As far as the representation theory of
the $\widehat{su}(2)$ affine Lie algebra is concerned, the $\zet_2$ 
automorphism $g$ is clearly realized in the untwisted (\ref{twin}) as well
as in the twisted sectors, (\ref{twist}), (\ref{twisttwin}). However, 
as can be seen for instance from the explicit fusion rules \cite{bifs},
$g$ can not be a symmetry of the full OPE. We have already seen that
$g$ squares to $(-1)^l$ even in the untwisted sector, where it generates 
$G_c \cong \zet_4$. Including the twist, which is measured by the quantum
symmetry group $G_q\cong \zet_4$, we expect that in general the OPE will
have a symmetry group $\Gamma$ which is an extension
\begin{equation}
G_c\cong \zet_4 \rightarrow \Gamma \rightarrow \zet_4 \cong G_q \,,
\end{equation}
depending on the value of $k\bmod 4$. Specifically, we expect
$\Gamma\cong\zet_{16}$ for $k$ odd, $\Gamma\cong\zet_2\times \zet_8$ for 
$k=2\bmod 4$ and $\Gamma\cong\zet_4\times\zet_4$ for $k=0\bmod 4$.
Only in the last case would an orbifold by $g$ make sense.

In any event, as is by now familiar, even if we cannot orbifold by 
$g$, we can still wilsonize it by fibering the sectors
(\ref{blocks}) over a free boson theory in the appropriate way.
To this end, we will use again the path-integral
sectors with fixed winding around space- and time-like worldsheet
circles introduced in (\ref{inverse}), (\ref{Ys}). The partition 
function of the Wilson line theory is then given by
\begin{equation}
Z = \sum_{a,b=0}^3\sum_{m,n\in\zet} \boi_a^b Y^{4n+ b}_{4m+a}
\zeta^{4n a - 4m b + b \delta(a)}
\label{belief}
\end{equation}
where the important phase corrections are given by
\begin{equation}
\delta(0)=0\qquad \delta(1)=1\qquad\delta(2) = -2 \qquad
\delta(3)=-1
\end{equation}
It is not hard to check that this expression is modular invariant,
and has a $q$-expansion with positive integer coefficients. Specifically,
we find that the projection in the $a$-th twisted sectors is onto 
states with momentum around the base $p\in\frac14\zet- \frac{k a}{16}$, 
and eigenvalue of $g$ in the fiber theory given by
\begin{equation}
\omega = e^{-2\pi i [ p - k m/4+ \delta(a) k/16]}
\end{equation}
where $m$ is related to the winding $w$ around the base circle 
(radius $R$) by $w=4m+a$. 

This clearly shows that on the one hand, the projection in the fiber
theory for unwound strings on the base is the correct one for a Wilson
line theory implementing charge conjugation in the fiber around the 
base circle. On the other hand, the momentum in the twisted, wound 
sectors is generally fractionated in units of 
$1/16=1/4^2$. Only for $k$ divisible by $4$ does this Wilson line 
theory fall into the ``tame'' category.

\section{General Theory of Wilson Line CFT}\label{general}

We have seen that Wilson line
theories for discrete gauge symmetries in
string theory exist in all the examples we examine.
Two striking features are apparent from
the set of examples we have considered so far.
First, the momentum in the base is not always fractionated
in the units we expect, namely ${1\over{N R\ll{base}}}$ 
where $N$ is the order of the symmetry group
of the fiber theory ${\cal I}$.  Secondly,
it is possible to wilsonize a symmetry even when one
cannot consistently orbifold by it.

\subsection{General properties of states and operators}

It is now worth deriving some general properties of
Wilson line CFTs in order
to shed light on the relationship between these
two observations.  The goal of this section will be to formulate
a criterion distinguishing those Wilson line 
CFT which can be represented straightforwardly
as $\IZ\ll N$ orbifolds from those which cannot.
We begin by deriving some properties which hold
for Wilson line CFT in general. To do this, let us formalize our
definition of a Wilson line in string theory.  We will take
the base to be $S\uu 1$, though the generalization to a $T\uu k$ base
is self-evident.
\vsk
${\bf{\underline{Def:}}}$ A Wilson line theory a
CFT with operator algebra ${\cal W}$, whose data are a fiber
CFT ${\cal I}$, a CFT base $S\uu 1$ 
and an anomaly-free
discrete symmetry $g$ of finite order $N$ acting on ${\cal I}$.
The operator algebra of the base is a $U(1)\ll L
\times U(1)\ll R$ current algebra ${\cal A}\uu{S\uu 1}$ with
a Narain lattice $\L = \{(p\ll L\equiv
p\ll{base} - {1\over{\apr}} R\ll{base}
 w\ll{base} , p\ll R\equiv p\ll{base} + {1\over{\apr}} R\ll{base}
w\ll{base}) \}$ which is neither necessarily
even nor self dual.  The Wilson line CFT ${\cal W}$
has the following defining properties:
\begin{indent}
\bi
\item {The subalgebra of ${\cal W}$ with $w\ll{base}
= 0$ is the subalgebra of ${\cal A}\uu{S\uu 1} \otimes {\cal I}$
such that $g$ and $g\ll X\uu{-1}
\equiv \exp{-2\pi i R\ll{base} p\ll{base}}$ have the
same eigenvalue.}
\item{${\cal W}$ has all the standard properties
of a well-behaved CFT, in particular modular invariance, and
single-valuedness and closure of the operator product expansion.}
\ei
\end{indent}
\vsk
${\bf{\underline{Property~1:}}}$ For all Wilson line theories, the
worldsheet boundary conditions on operators in ${\cal I}$ are
a function of the winding $w\ll{base}$.  In particular, operators in
${\cal I}$ have a periodicity ${\cal O}\to g\uu {w\ll{base}}
 \cdot {\cal O}$ around
the spatial coordinate of the worldsheet in the sector with winding
$w\ll{base}$. 
\begin{indent}
\vsk
\it Proof: \rm Consider an operator $V\ll 1$
with $w\ll{base} = 0$ and nonzero momentum 
$p\ll{base}\equiv  n\ll{base} /  R\ll{base}$.
We would like to transport $V\ll 1$ around the origin,
where we have placed an operator with winding $w\pr\ll{base}$.
When transported around $V\ll 2$, the operator
$V\ll 1$ gets a phase $\exp{2\pi i
w\pr \ll{base} n\ll{base}}$
from its base part.
The phase contributed by the fiber part of the operators
must cancel this phase.  In particular, if the boundary condition 
at $z = 0$ is
${\cal O} \to g\uu m \cdot {\cal O}$ and the $g$-eigenvalue
of $V\ll 1$ is $\o$, then $\o\uu m$ must equal
$\exp{2\pi i n\ll{base} w}$ for all $n\ll{base}$.  By the defining
property of a Wilson line CFT,
$\o = \exp{w\pi i n\ll{base}}$, so by
taking $n\ll{base} = {1\over N}$, we see that $m = w$ mod N.  But the
number $m$ is only meaningful mod N, since $g\uu N$
acts trivially on all operators in ${\cal I}$.  So the boundary condition
in the sector with winding $w\ll{base}$
is ${\cal O}\to g\uu {w\ll{base}} \cdot {\cal O}$, \it QED. \rm

\end{indent}
\vsk
${\bf{\underline{Property~2:}}}$ For all Wilson line theories, the
quantity $n\ll{base}\equiv R\ll{base} p\ll{base}$ is completely
determined mod 1 by $w\ll{base}$ and by the eigenvalue of $g$.

\begin{indent}
\vsk
\it Proof: \rm Consider two states $V\up 1$ and $V\up 2$ with the same
winding number $w\ll{base}$ and the same eigenvalue $\o$ of $g$.
Then $V\ll 2\dag$ has winding number $- w\ll{base}$ and $g$-eigenvalue
$\o\st$.  Then the OPE of $V\ll 1$ with $V\dag\ll 2$ contains states
with zero $w\ll {base} = 0$ and $g = 1$, which by the defining property
must have $n\ll{base} \in \IZ$.  Therefore $n\ll{base}$ values of 
$V\up 1$ and $V\up 2$ must differ by an integer, \it QED. \rm

\end{indent}

\heading{Wild versus tame Wilson lines\protect\footnote{We are very grateful
to A. Flournoy, B. Williams and B. Wecht for discussions relevant to
this section.}}

At this point we wish to draw a sharp distinction between two different
cases in which the symmetry $g$ is of order $N$ in its
action on all local fields in the base.

In the first case, which we shall refer to as 'tame', the level mismatch in
the fiber theory
in the sector twisted by $g$ is an integer multiple of
${1\over N}$.
This means that the momentum in the base is fractionated in units
of $p\ll{base} = {1\over{NR\ll{base}}}$.
Two properties of tame Wilson lines follow directly from this.
\vsk
$\underline{{\bf Property~3}}$: A tame Wilson line 
theory can be viewed as an orbifold of a product of the fiber
with an $N$-fold cover of the base.  The orbifold action
is $g$ on the fiber and a shift $X\ll{base} \to X\ll{base} + 2\pi N R\ll{base}$
on the base.  
Each state, twisted or untwisted, carries a representation of $g$ which
may be nontrivial, but nonetheless satisfies
the group relation $g\uu N = 1$ obeyed by $g$.
\vsk
$Proof:$
We can always change the momentum of any state mod ${1\over{NR\ll{base}}}$
without changing the winding or twist, by taking its
OPE with unwound states. 
By definition of a tame Wilson line
all states, including winding sectors,
carry momentum which vanishes mod ${1\over{NR\ll{base}}}$.
It follows that every winding sector
contains some states with $p\ll{base} = 0$.
In particular, the sector
with $w\ll{base} = 1$ contains a state with $p\ll{base} = 0$.

Take the OPE of $N$ copies of that state,
which produces a string with $w\ll{base} = N$ and $p\ll{base} = 0$.
Call this vertex
operator $V\ll 1$, and call its $g$-eigenvalue $\o\up 1$.
Now, consider what happens when we circle $V\ll 1$ around an
operator $V\ll 2$ in the twisted
sector which has $p\ll{base} = {n\over {N R\ll{base}}}$ ($n\in\IZ$).
The phase one gets
in circling $V\ll 1$ once around $V\ll 2$ is equal to
\bbb
\exp{2\pi i (n\up 1 w\up 2 - n\up 2 w\up 1)}~\cdot~
(\o\up 1)\uu{w\ll 2} ~(\o\up 2)\uu{- w\ll 1},
\eee
with the second and third factors coming from the fact that
the sector with winding $w\ll{base}$ is also twisted by $g\uu 
{w\ll{base}}$.
For the two vertex operators we are considering, the
first factor makes no contribution, since $n\up 1 = 0$ and
$n\up 2 w\up 1$ is an integer.  The third factor also makes
no contribution since $w\ll 1 = N$ and $\o\up 2$ is an $N\uu{{\rm \underline{th}}}$
root of
unity.  So the phase is just $\o\up 1$, since $w\ll 2 = 1$.
Now, for consistency of the OPE, the total phase must equal 1, so
$\o\up 1 = 1$, proving that all operators with $w\ll{base} \in N\IZ$
and $n\ll{base} \in \IZ$ must have $\o = 1$, \it QED. \rm

\vsk
\it Corollary: \rm By closure of the OPE it follows that $\exp{2\pi i 
n\ll{base}} = \a\uu {w\ll{base}}
 \cdot \o$, with $\a$ an N$\uu{\rm{\underline{th}}}$
root of unity.  But
since our symmetry $g$ was originally defined only in ${\cal I}$,
we can choose to extend it to ${\cal I} \cdot {\cal A}\uu{S\uu 1}$ in
any way which respects the group relation $g\uu N = 1$.  In particular
if $\a$ is a root of unity we can redefine $g\to \a\uu{-w}\cdot g$, in
which case our formula becomes $\exp{2\pi i 
n\ll{base}} = \o$, for all states in the theory.

\vsk

$\underline{{\bf Property~4}}:$
The $R\ll{base}\to 0$ limit of a tame Wilson line
is the product of an orbifold ${\cal I} / \{g\}$
of the fiber with the real line $\IR$.  The real line comes
from the decompactification of the T-dual coordinate $\tilde{X}\ll{base}$
to the base $X\ll{base}$.

\vsk

$Proof:$
The structure of the $R\ll{base} \to 0$ limit can be understood
from consideration of the states which become light in the limit $R\ll{base}
\to\infty$, which are exactly the states with $p\ll{base} = 0$.
Then the states
which survive in the limit $R\ll{base}\to 0$ must have $n\ll{base} = 0$,
which implies $\o = 1$.  And for every winding state, there is
a sector whose boundary conditions are ${\cal O}\to g\uu {w\ll{base}}
\cdot {\cal O}$ around the spatial coordinate of the worldsheet.

So the theory at large $R\ll{base}$ describes a $g$-invariant set of
states in ${\cal I}$ which form a continuum with density
${{d{\cal N}}\over{d\tilde{p}\ll{base}}} = \tilde{R}\ll{base}$,
where
\bbb
\tilde{p}\ll{base} = \tilde{R}\ll{base}
\\\\
\tilde{R}\ll{base} = {{\apr}\over{R\ll{base}}}
\eee
and a sector twisted by each element of $G$, with the same
spectral density ${\cal N}$.
It follows that the $R\ll{base} \to 0$ limit
of a tame Wilson line is simply the product of $\IR\ll{\widetilde{base}}$
with the orbifold ${\cal I} / G$, \it QED. \rm

\heading{The wild case}

In the more general case, the momentum $p\ll{base}$ does not satisfy
$N n\ll{base} \in \IZ$ for all states.  However we will now show
that the denominator of $n\ll{base}$ does obey a general bound.

\vsk

\bf \underline{Property 5}: \rm The momentum $n\ll{base}$ always has
a denominator dividing $N\sqd$.
That is, $N\sqd n\ll{base} \in \IZ$ for
any consistent Wilson line CFT. 

\vsk

\it Proof: \rm It suffices to prove that the momentum in the
sector with $w\ll{base} = 1$ satisfies $N\sqd n\ll{base} \in \IZ$.
Let $n$ be the value of $n\ll{base}$ in any sector with $w = 1$.
Then take the OPE of $N$ copies of the state.  The resulting piece has
an untwisted fiber factor, so its contribution to the level mismatch
is $0$ mod 1.  Therefore the base contribution to the level mismatch
is also $0$ mod 1.  But the base contribution is $N \cdot (N n\ll{base})$,
which means the $n\ll{base}$ we started with in the $w\ll{base} = 1$
sector must be an integer multiple of ${1\over{N\sqd}}, QED.$  

It follows that the value of $n\ll{base}$ in the sector with $w\ll{base} = N$
is an integer multiple of ${1\over N}$, so there does exist a
state $V$ with winding $N$ and $p\ll{base} = 0$.  In general
however, this state will not have $\o = 1$.


\vsk

$\underline{\bf Property~6}:$
The value of $p\ll{base}$ mod ${1\over{NR\ll{base}}}$ is
completely determined by the winding:
\bbb
n\ll{base} = 
{{\a\ll 1 w\ll{base}}\over{N\sqd}} \llsk\llsk\llsk\rm (mod {1\over
{\it N \rm}}) 
\eee
where $\a\ll 1$ is some integer characteristic of the theory.

\vsk

$Proof:$ For a state with
$w\ll{base} = 1$ we have $n\ll{base} =
{{\a\ll 1}
\over{N\sqd}}$ (mod 1) by property 5, so \it a fortiori \rm 
we have $n\ll{base} =
{{\a\ll 1}\over{N\sqd}}$ (mod ${1\over N}$).  The OPE of 
$b$ copies of state yields a state $V\up 1$ with $w\ll{base} = b$ and
$n\ll{base} = {{\a\ll 1 b}\over{N\sqd}}$ (mod ${1\over N}$).
If another state $V\up{2}$ also has $w\ll{base} = b$, then
$V\up 1{}\dag V\up 2$ has $w\ll{base} = 0$.  From the
defining property of a Wilson line CFT and the
conservation of $p\ll{base}$ in the OPE, it follows that
$V\up 2$ has $n\ll{base}
= {{\a\ll 1 b}\over {N\sqd}}$ (mod ${1\over N}$), $QED$.

\vsk 

$\underline{\bf Property~7}:$
In a general winding sector, the eigenvalue $\o$
of $g$ is always an $
\lrdd N\sqd\rrdd \uu{\rm{\underline{th}}}$
root of unity.

\vsk

$Proof:$ Let the winding $w\ll{base}$ of the state be $b$
and its $g$-eigenvalue be $\o$.  The OPE of $N$ copies
of the state has $w\ll{base} = N b$ and $g$-eigenvalue $\o\uu N$.
Since $w\ll{base} \in N\IZ$, the fiber component of
the state is untwisted, by property 1, which means
it lives in ${\cal I}$.  The $g$-eigenvalues of
all states in ${\cal I}$ are $N\uu{\rm{\underline{th}}}$
roots of unity, which means $(\o\uu N)\uu N = \o\uu{N\sqd} = 1$.
In other words, $\o$ is an $
\lrdd N\sqd\rrdd \uu{\rm{\underline{th}}}$
root of unity, $QED$.

\vsk

$\underline{\bf Property~8}:$
The eigenvalue $\o$ of $g$ 
mod $\exp{2\pi i / N}$ is completely determined by
the winding:
\bbb
\o = \exp{2\pi i \b\ll 1 w\ll{base}/ N\sqd}\llsk\llsk\llsk\rm (mod~
\exp{2\pi i / \it N \rm} )
\eee
where $\b\ll 1$ is some integer characteristic of the theory.

\vsk

$Proof:$ For a state with $w\ll{base} = 1$, we have 
$\o = \exp{2\pi i \b\ll 1 / N\sqd}$ with
$\b\ll 1\in \IZ$ by property $7$.  So
\it a fortiori \rm we have $\o = \exp{2\pi i \b\ll 1 / N\sqd}$
(mod$~
\exp{2\pi i /  N } $).  Taking the OPE of $b$ copies
of such a state yields a state $V\up 1$ with $w\ll{base}
= b$ and $\o = \exp{2\pi i \b\ll 1 b / N\sqd}$
(mod$~
\exp{2\pi i /  N } $).  
If another state $V\up{2}$ also has $w\ll{base} = b$, then
$V\up 1{}\dag V\up 2$ has $w\ll{base} = 0$.  From the
defining property of a Wilson line CFT and the
conservation of $g$ in the OPE, it follows that
$V\up 2$ has $g$-eigenvalue $\o$ equal to
$\exp{{{2\pi i \b\ll 1 b}\over {N\sqd}}}$ (mod
$\exp{2\pi i /  N } $), $QED$.

\vsk

$\underline{\bf Property~9}:$ The product
$\phi\uu{-1}\equiv \o \cdot \exp{2\pi i n\ll{base}}$, where $\o$ is
the eigenvalue of $g$, depends only on the value of
$w\ll{base}$.

\vsk

$Proof:$ Let $V\up{1,2}$ be two states with the same
values of $w\ll{base}$.  Take the OPE of
$V\up 1$ with $V\up 2{}\dag$.  The quantum numbers
$n\ll{base}$ and $g$ are exactly conserved
by the OPE, so the product $\phi\uu{-1}
\equiv \exp{2\pi i n\ll{base}}
\cdot \o$ is conserved as well.  The product $V\up 1
~V\up 2 {}\dag$ has $w\ll{base} = 0$, so by the
defining property of a Wilson line CFT it has
$\exp{2\pi i n\ll{base}}
\cdot \o = 1$.  It follows that $V\up 1$ and $V\up 2$
share the same value of $\phum = \exp{2\pi i n\ll{base}}
\cdot \o , QED$.

\vsk

$\underline{\bf Property~10}:$
Given a Wilson line CFT, there is a formula which applies
to all physical states in the spectrum:
\bbb
\exp{2\pi i n\ll{base}}~\o = \phi\ll 1 \uu {-w\ll{base}} 
\een{genform}
where $\phi\ll 1$ is an $\lrdd N\sqd\rrdd \uu{\rm{\underline{th}}}$
root of unity.

\vsk

$Proof:$ Given some state in the sector
$w\ll{base} = 1$ define $\phi\ll 1\uu{-1}$
to be the product $\exp{2\pi i n\ll{base}}
\cdot \o $.  Both $n\ll{base}$ and $g$ are
$\lrdd N\sqd\rrdd \uu{\rm{\underline{th}}}$
are roots of unity in the sector $w\ll{base} = 1$, so
$\phi\ll 1$ is an
$\lrdd N\sqd\rrdd \uu{\rm{\underline{th}}}$ root
of unity as well.  Taking the OPE of $b$ copies of this state
yields a state with $w\ll{base} = b$ and
$\o \cdot \exp{2\pi i n\ll{base}} = \phi\ll 1\uu {-b}$.  By
property $9$, $\o\cdot \exp{2\pi i n\ll{base}}$ depends
only on $b$, so $\o \cdot \exp{2\pi i n\ll{base}} = \phi\ll 1\uu 
{-b}$
holds for all states.
The formula (\ref{genform}) follows, $QED.$

\vsk

\it Corollary: \rm The orbifold-like projection
$\o = \exp{-2\pi i n\ll{base}}$ cannot
be extended in general from the $w\ll{base} = 0$
sector to the winding sectors, even if
$g$ is extended to generate a $\IZ\ll{N\sqd}$ group
when acting on twisted sectors.
The
action of $g$ on untwisted sectors is already
defined on operators in ${\cal I}$, which are
the internal pieces of vertex operators with
$w\ll{base} \in N\IZ$.  And with that prior definition of
$g$, the projection $\o = \exp{-2\pi i n\ll{base}}$
is not even obeyed in sectors with $w\ll{base} \in N\IZ$.  

\heading{Light states in wild Wilson theories in the $R\ll{base}
\to 0$ limit}

What are the light states, in this case, as $R\ll{base}\to 0$?  They 
are still the states with $p\ll{base} = 0$.  But now the
set of such states
have fiber pieces which get a nontrivial phases under 
some elements $g$.  This 
establishes that the $R\ll{base}\to 0$ limit does \it
not \rm contain a
factor of ${\cal I} / G$.

To see what we do get in the limit $R\ll{base}\to \infty$, consider
the character $\chi$ defined by the phase $\chi(g) = \o$, with
$\o$ taken to be the eigenvalue of $g$ in the sector with $w\ll{base} = N$
and $p\ll{base} = 0$.  Suppose $\chi$ generates the entire dual group
$G\st$ of $G$.  This means that there is no nontrivial element
$h$ of $G$ 
such that $\chi(h) = 1$.  Concretely, it means $\chi(g) = \exp{2\pi i
m  / N}$ with $m$ relatively prime to $N$.  In this case, all $N$
possible eigenvalues of $g$ occur for various states with
winding $w\ll{base}$ in $N\IZ$ and $p\ll{base} = 0$.  In such
a case, the first state with $p\ll{base} = 0$ and $\o = 1$
occurs for winding $w\ll{base} = N\sqd$.  As $R\ll{base}\to 0$,
these states form continua in a
T-dual theory at large radius, with spectral density
\bbb
{{d {\cal N}}\over{d\tilde{p}}} = \tilde{R}\ll{base}
\\\\
\tilde{R}\ll{base} = {{N\sqd \apr}\over{R\ll{base}}}
\eee
for each sector.  The twisted sectors have $n\ll{base} \notin {1\over N}
\IZ$ so in particular they never have vanishing momentum, and thus
they become infinitely heavy in the limit $R\ll{base} \to 0$.

The resulting theory, in the 'maximally wild' case gcd$(m,N)
= 1$, has the
full complement of states in ${\cal I}$, with all possible
eigenvalues of $g$, and no twisted sectors for small $R\ll{base}$.
It follows that the limit $R\ll{base} \to 0$ of the 'maximally wild'
Wilson line is simply $\IR\ll{\widetilde{base}} \times {\cal I}$.
Intermediate cases can occur as well, in which
the denominator of the momentum fractionation
is an integer multiple of $N$ dividing $N\sqd$.

\subsection{Effective field theory on the long winding string}

The wild momentum fractionation in units
of ${1\over{N\sqd R\ll{base}}}$ may at first seem exotic
and its introduction into the CFT \it ad hoc. \rm 
One might be tempted to object that effective field
theory would forbid, rather than mandate, the inclusion of states
with $p\ll{base} \notin {1\over N} \IZ$.

The point is, this intuition is correct insofar as it refers to effective
field theory in the bulk of the base in the limit $R\ll{base} >> \sqrt{\apr}$.
Bulk states -- states which are local fields in the base -- do
indeed have $p\ll{base} \in {1\over N} \IZ$ with values mod $1$ according
to their $g$-eigenvalues, as dictated by the defining property of a Wilson
line.  But in the same limit one can analyze the effective theory on
a long string winding the base, with attention to the spectrum of
momenta $p\ll 9$ for states of the long string.

What is the effective field
theory on a long string with winding $w\ll{base} = 1$?
Since our backgrounds are represented by conventional
worldsheet CFT with no fluxes or warping, the energies are given by
\bbb
E =
  \lrdd p\ll i p\ll i + p\ll {base}\sqd 
+
{{R\ll{base} \sqd w\ll{base}\sqd}\over{\apr}} +
{2\over{\apr}}
H\ll{\perp}\uu{w.s.}\rrdd\uu\hh
\eee 
where $p\ll i$ are the momenta in all noncompact
spacelike, and $m\sqd\ll{ \perp}$ is the mass-squared ${1\over{\apr}}
H\uu
{w.s.}\ll{ \perp}$ contributed by the component of the
worldsheet Hamiltonian $H\uu{w.s.}$
other than $X\ll{base}$ and $X\uu 0$
(and their superpartners, in the heterotic or type II theory).
So at long wavelengths in the c.m. frame $p\ll i = 0$ we have
\bbb
E\sim {{R\ll{base}}\over{\apr}} +  
{1\over{R\ll{base}}}
H\ll{\perp}\uu{w.s.} 
\eee

Since $H\uu{w.s.}\ll{\perp}$ is a 
Hamiltonian with respect to worldsheet time, it
is dimensionless in the spacetime
sense.  For $R\ll{base} >> \sqrt{\apr}$, 
the theory on the long string
is approximately conformal, with Hamiltonian
$ {1\over{R\ll{base}}} H\uu{w.s.}$ plus a constant.
(The term $p\ll{base}\sqd$ is proportional to
two powers of $\sqrt{\apr} / R\ll{base}$ and is
therefore subleading in the limit of interest.)
The Hamiltonian of the \it long-wavelength effective CFT \rm
of the string wound on the base is therefore proportional
to the Hamiltonian of the \it fundamental string CFT \rm 
for the string degrees of freedom other than
$X\uu 0, X\ll{base}$, with the worldsheet
coordinates $(\s\uu 1, \s\uu 2)$ replaced by ${1\over{R\ll{WS}}}
(X\uu 0, X\ll{base})$.

This much is common to winding string effective field
theory in any compactification with a large $S\uu 1$
direction.  However the presence of the Wilson
line means that the boundary conditions on
the long string are 'twisted' by the
identification between the long string
worldvolume degree of freedom and the
base direction.  Treating $\s\uu 1 \equiv X\ll{base}/
R\ll{base} $ as an
independent worldvolume coordinate and
letting ${\cal O}$ be any operator in ${\cal I}$,
the boundary condition for ${\cal O}$ in the
long string theory must be
\bbb
{\cal O}(y\ll 1 + 2\pi ) = g\cdot {\cal O}(y\ll 1)
\eee
in order for effective field theory on the long string
to be consistent with effective field theory for
the bulk modes in the base.

In other words, the long string Hilbert space is
actually described by a 'twisted sector' of the
internal theory ${\cal I}$.  Such a sector would
not be consistent as a state in a modular invariant
worldsheet theory, but the long string effective theory
need not satisfy modular invariance -- it is impossible
to perform a Dehn twist without changing the topology
of the string in the target space, so there is no
reason for modular invariance to be a good symmetry in
this theory.

However the physical consequence of the failure of
modular invariance is important.  The sector twisted by $g$
will in generally fail to satisfy level matching, by
an amount $\tilde{L}\ll 0 - L\ll 0 = a$ (mod 1).  This
quantity is a ground state value for the
coordinate momentum $P\ll{\s\uu 1}$ whose value can
only be raised and lowered in units of ${1\over N}$ 
by modes of local operators on the long string worldsheet.
As we know from free field examples, the level mismatch
in the sector twisted by $g$ will in general lie in
${1\over{N\sqd}} \IZ$ rather than ${1\over N} \IZ$.
Using the coordinate identification $\s\uu 1 = {1\over {R\ll{base}}}
X\ll{base}$, we find that the level mismatch translates into
a ground state value of the spacetime momentum
in the base given by
\bbb
p\ll{base} = {{\theta}\over{R\ll{base}}}
\eee
where $\theta$ in general lies in ${1\over{N\sqd}}\IZ$ and not
in ${1\over{N}}\IZ$, as we concluded from considerations
of modular invariance in the fundamental string CFT.

So we find that the ${1\over{N\sqd}}$ momentum fractionation
in the base of the wild Wilson line string theory is not
a mysterious worldsheet artifact but an inevitable consequence
of the consistency of the effective dynamics of string
theory at wavelengths long compared to the string scale.

\subsection{Are flat monodrofolds orbifolds?}

We should make a brief comment about the relation
between the standard 'orbifold' construction
and the construction of stringy Wilson lines 
described in this section.  Are stringy
Wilson lines just orbifolds of some
kind?  And if they are, what kind of orbifolds
are they?

In terms of a projection on the 
sector with $w\ll{base} = 0$ is
identical to the construction of the untwisted
sector of an orbifold of ${\cal I} \times S\uu 1$,
where the radius of the $S\uu 1$ is $2 N\pi R\ll{base}$.
The projection in the sector $w\ll{base} = 0$
is $g\cdot g\ll X = 1$, where $g$
acts on the fiber and $g\ll X$ acts by
$X\ll{base} \to X\ll{base} + 2\pi R\ll{base}$.

Modular invariance also demands the inclusion of
twisted sectors, where the sector with winding $w\ll {base}$
has $g$-covariant fields twisted by $g\uu{w\ll{base}}$.
However the relation between the eigenvalue $\o$ of
$g$ and the momentum $R\ll{base} p\ll{base}$ mod 1
is not the same in the winding sectors as the
naive orbifold projection would dictate.  This can be
fixed by including a winding-dependent phase
in the orbifold projection in the winding sectors:
\bbb
g \cdot g\ll X = 1 \to g\cdot
g\ll X = \O\ll{w\ll{base}}
\eee

Broadly speaking this can be considered a type of
orbifold projection.  But one must use some caution,
because the most straightforward application of the
orbifold idea leads to incorrect conclusions.
The orbifold projector is the generator
of a group with structure $\IZ\ll N$ on the covering space.
That is, both $g\uu N$ and $g\ll X
\uu N$ act separately
as the identity
all states on $S\uu 1\ll{cover}\times {\cal I}$.
However neither $g$ nor $g\ll X$ nor
the product of the two is
of order $N$ when extended to the twisted sector.
In particular, if $g\ll X$ were of order $N$,
all states would have momentum fractionated
in units of ${1\over{N R\ll{base}}}$.  Likewise if
$g$ were of order $N$ when extended to the twisted sector,
the OPE of $N$ identical states would always be $g$-neutral.
Neither is the case in general.
So although one may refer loosely to
Wilson lines of wild type as orbifolds of an $N$-fold cover, it
is good to bear in mind that the simple terminology
conceals
important features of the spectrum.

The partition function which implements such a projection
is of the form
\bbb
Z\equiv \sum\ll{a,b\in \IZ} Y\uu a\ll b I\uu a\ll b
\phi\ll 1\uu{ab}
\\\\
= \sum\ll{a,b\in \IZ} Y\uu a\ll b I\pr{}\uu a\ll b
\eee
where $I\uu a\ll b$ is the trace over states of the internal
theory (here, all degrees of freedom other than $X\ll{base}$)
in the sector twisted by $g\uu b$, with $g\uu a$ inserted
into the trace.  The $I\pr{}\uu a\ll b \equiv \phi\ll 1\uu{ab}
I\uu a\ll b$
are the related objects which transform classically
under modular transformations.  The phase $\phi\ll 1$
is obtained by exponentiating the level mismatch
\bbb
\lno\hilo
\phi\ll 1 \equiv \exp{2\pi i (\tilde{L}\ll 0 - L\ll 0)}
\rba\ll{\rm [sector~of~{\cal I}~\rm twisted~by~\it g\uu 1\rm]}
\eee
evaluated in the sector of the internal theory twisted
by $g\uu 1$.  All the examples we consider in the
paper obey this classification.\footnote{Our $SU(2)$
current algebra
examples at level $k$ use a different basis
of internal partition functions, $\boi\uu a\ll b$, 
which are related to the $I\uu a\ll b$ by phases.}

\section{A monodrofold of the type II superstring}\label{typeii}

The next most nontrivial case is a monodromy by a chiral action
$g$ which acts on moduli as $\t\to - {1\over \t}, \rho\to - {1\over\rho}$.  The fiber theory ${\cal I}$ is
a square torus $X\uu{89}$, 
with each circle at the self-dual radius, and also worldsheet
fermions $\psi\uu{8,9},\pst\uu{8,9}$.
The operation $g$ which sends $\t\to - {1\over \t}$ and $\r \to - {1\over\r}$
acts on
left-moving worldsheet currents as
$J\uu{8,9}\ll - \to - J\uu{8,9}\ll -$.
By worldsheet supersymmetry it must also act on left-moving
worldsheet fermions as
$\pst\uu{8,9}\to - \pst\uu{8,9}$.  

The discussion of the action of T-duality on
the bosons $X\uu{8,9}$, as well as the corresponding
partition functions and construction of the
twisted sectors, is all precisely as it is
for a single boson $X\uu 9$.  
In particular, all the discussion
of cocycles and zero-mode dependent
phases is the same for two copies of a self-dual circle
as it is for one.

In addition to acting with a $-$ sign on
left-moving oscillators in the $8,9$ directions, the
action of T-duality
it also acts with a phase $(-1)\uu{n\ll 8 w\ll 8 + n\ll 9 w\ll 9}$,
where $n\ll {8,9}$ and $w\ll{8,9}$ are the windings and momenta of
the state on the torus.  The argument is precisely as in the
case of the bosonic string; the T-duality operation which is
multiplicatively conserved in the OPE is the one with the extra
cocycle factor in its action on momentum and winding.  Since the symmetry
acts as a 180 degree rotation on the left handed
89 currents, it must act as a 90 degree phase rotation
$\exp{i \G\uu 8 \G\uu 9} = \exp{i \pst\ll 0 \uu 8 \pst\ll 0\uu 9}$
on left-handed
R sectors.  The symmetry is therefore $\IZ\ll 4$ rather than $\IZ\ll 2$ in
its action on spacetime fields.

The partition functions $I\up {X\uu 8}{}\uu a\ll b$ and
$I\up {X\uu 9}{}\uu a\ll b$ are the same as 
the partition functions $I\uu a\ll b$ 
for a single $S\uu 1$ fiber of self-dual radius in the bosonic
string.  The $I{}\uu{(X\uu 8, X\uu 9)}{}\uu a\ll b$ for
the bosonic piece of the torus are just the square
of the $I\uu a\ll b$ for a single self-dual circle.  This
is the fact expressed in $(\ref{morefibers})$, with $k = 2$.

The Wilson line for the $\IZ\ll 2$ T-duality in the
bosonic string
is tame, so each partition function $I\up {X\uu{8,9}}{}\uu a
\ll b$ has classical transformations under $PSL(2,\IZ)$.
The only new element in this section is the discussion
of the worldsheet fermions, which will change
the symmetry in the untwisted sector to $\IZ\ll 4$
and also make the Wilson line for T-duality in the
type II string wild rather than tame.

\subsection{Worldsheet fermions in the type II superstring}

In the type II superstring the
transformation laws of
the internal pieces are not as
simple.
In this case, it will be convenient to
include all worldsheet fermions $\psi\uu\m, \pst\uu\m$
and superghosts $\b,\tilde{\b}, \g, \tilde{\g}$ in
the partition function as well.  Alternately we can omit
the superghosts and the fermions $\psi\uu{0,1}, \pst\uu{0,1}$;
the result is the same.

The partition functions $\fft\uu a\ll b(\tb)
\equiv \qb\uu{- {1\over{6}}}
\tr\ll{\tilde{\b},\tilde{\g},\tilde{\psi},
[\rm twisted~by~\it g\uu b]} 
\lrdd
\qb\uu{\tilde{L}\ll 0} g\uu a \rrdd
$ for the left-moving fermions and superghosts
are described by the following properties. 
\bi
\item{Each is a sum of four terms
$\fft\uu a\ll b \equiv \hh 
\o\uu {a|c}\ll{b|d}
(\bfft\uu c\ll d)\uu 3 \bfft\uu {a+c}\ll{b+d}$ with
$\o\uu {a|0}\ll{b|0} = i\uu{ab}$ and $|\o\uu {a|c}\ll{b|d}| = 1$,
with the $\bfft$ defined in (\ref{complexfermionpartitionfunctions}).
The right-hand indices denote boundary conditions for
the supercurrent and the left-hand indices denote
boundary conditions for $g$-covariant fields.}
\item{These phases satisfy $\o\uu{a+2p|c}\ll{b+2q|d}
 = (-1)\uu{p(b+d)+q(a+c)} \o\uu{a|c}\ll{b|d}$.  That is,
shifting $a$ by two changes the sign of the untwisted Ramond
sectors and twisted NS sectors
in the path integral, and shifting $b$
by two changes the sign of the GSO and orbifold projections.  In
particular, the $\o$'s, and hence the $\fft\uu a\ll b$,
are periodic mod four in each index.  All the
$\fft$ are determined in this way by $\fft\uu a\ll b$ with
$a$ and $b$ running from $0$ to $1$, which
we write below.}
\item{ The modular $T$-transformations of the $\fft$ are
determined only by their
lower indices: $\fft\uu a\ll b ( \t + 1) =
\exp{{{2\pi i}\over 3}}~\O\ll b~\fft\uu {a+b}\ll b (\t)$
with $\O\ll 0 = 1, \O\ll 1 = \O\ll 3 =
\s \exp{- {{\pi i}\over 4}}
,$ and
$\O\ll 2 = -1$.}   
\item{For the definition in terms of traces to make
sense, we must define $g\sqd$ to act as $(-1)$ on left-moving
Ramond sectors, which is implied by closure of the
OPE: the product of two left-moving R sectors closes
on operators with an odd number of fermions $\psi\uu{8,9}$,
since
the total number of $g$-odd complex fermions
is 1.  This feature of the OPE can be
seen by bosonizing the fermion.  Then $S$-invariance
demands that we define 
the sector twisted by $g\sqd$ to be a sector with all
GSO projections reversed in sign relative to the usual GSO.}
\item{The modular $S$-transformations are $\fft\uu a\ll b( - 
{1\over \t}) = 
i\uu{-ab} \fft\uu{-b}\ll a (\t)$.  
}
\item{The anomalous phases $\O\ll b$ in the 
T-transformations encode the level mismatch mod 1
in the NS sector twisted by $g\uu b$, appropriately GSO projected.
In the
unwtwisted sector the level
mismatch is 0; in the first and third twisted
sectors it is ${3\over 8}$.  In the second twisted
sector the level mismatch is $\hh$; this is simply the reversal of
the GSO projection.  
These values assume $\s = -1$. 
 Taking
$\s = + 1$ simply changes the level mismatch to $- {1\over 8}$
instead of $+ {3\over 8}$ in the first and third 
twisted sectors.  Note that the level mismatch in
the third twisted sector is always the \it same \rm as 
that in the first twisted sector, rather than differing by
$\hh$.  That is because \it both \rm the GSO and orbifold
projections are opposite between the first and third twisted
sectors, so the difference can be made up by acting with a
zero mode $\pst\uu{8,9}\ll 0$.
}
\ei
The four functions which determine the
rest are
\bbb
\fft\uu 0\ll 0 (\tb) = \hh \lrdd \fft\uu {0|0} 
\ll{0|0} (\tb)
- \fft\uu{0|1}\ll{0|0}(\tb)
- \fft\uu{0|0}\ll{0|1}(\tb)
  \mp \fft\uu{0|1}\ll{0|1} (\tb)
\rrdd
\\\\
\fft\uu 1\ll 0 (\tb) =\hh \lrdd \fft\uu {1|0} 
\ll{0|0} (\tb)
- 
\fft\uu{1|1}\ll{0|0}(\tb) 
-i \fft\uu{1|0}\ll{0|1}(\tb)
 \mp i \s \fft\uu{1|1}\ll{0|1}(\tb) \rrdd
\\\\
\fft\uu 0\ll 1 (\tb) = \hh \lrdd \fft\uu {0|0} 
\ll{1|0} (\tb)
+ i \s
\fft\uu{0|1}\ll{1|0}(\tb) 
- \fft\uu{0|0}\ll{1|1}(\tb)
 \pm i \s \fft\uu{0|1}\ll{1|1}(\tb) \rrdd
\\\\
\fft\uu 1\ll 1 (\tb) =
{i\over 2} \lrdd \fft\uu{1|0}\ll{1|0} - i \s \fft\uu{1|1}
\ll{1|0}
- \s \fft\uu{1|0}\ll{1|1} \pm i \fft\uu{1|1}\ll{1|1} \rrdd
\eee

\subsection{The full type II theory}

Now we combine all the elements of
the type II worldsheet theory:

\bi
\item{The $\fft\uu a\ll b$ functions of the left-moving
superghosts and fermions.}
\item{The partition functions $Z\ll{\b\g \psi}
, Z\ll {bc\tilde{b}\tilde{c}},Z\ll{ X\uu{0-6}}$
for all the worldsheet fields
which do not depend on $a$ and $b$,
namely  the bosonic ghosts, worldsheet fields
$X\uu{0-6}$, and right-moving superghosts
and worldsheet fermions.  
(The latter are already defined as sums over right-moving
spin structures.)  These objects are inert under
modular transformations (except for the gravitational
anomaly $\exp{-{{2\pi i}\over 3}}$ in the 
T-transformation $\t \to \t + 1$ of $Z\ll{\b\g\psi}$
coming from the unbalanced
central charge.)}
\item{The partition functions $I\up{X\uu 8 X\uu 9}\uu a\ll b
\equiv (I\up{X\uu 9}\uu a\ll b)\sqd$
of the bosonic coordinates of the fiber torus.  As we
have seen, these path integrals have classical
modular transformation properties.  Under $S$ and $T$, the
indices $a$ and $b$, which run from $0$ to $1$, simply
transform as a doublet under $SL(2,\IZ)$, with
no anomalous phases in the $I\up{9}\uu a\ll b$.}
\item{The path integrals $Y\uu a\ll b$ for
the bosonic base coordinate $X\uu 7$, where $a$
and $b$ define the windings of the two cycles of
the worldsheet torus around the target space circle
$X\uu 7$.  These functions transform classically
under $SL(2,\IZ)$.}
\item{The partition functions $\fft\uu a\ll b$ of
the left-moving superpartners of $X\uu{8,9}$.  These objects
are periodic mod 4 in each index, and
they transform with anomalous phases.  That is, they define
a projective representation of $SL(2,\IZ / 4)$.}
\ei
We can then combine $Z\ll{\b\g\psi}$ with $\fft\uu a\ll b$
and and $I\uu{(X\uu 8 X\uu 9)}{}\uu a\ll b$
to get an object whose gravitational anomaly
cancels, and whose anomalous transformations
are entirely due to a failure of level matching.
We will call this full object $I\uu a\ll b$, and
we can equally well include the other factors
$Z\ll {X\uu {0-6}}, Z\ll{bc\tilde{b}\tilde{c}}$
with it.  Its anomalous transformation properties come
entirely from those of the $\fft\uu a\ll b$:
\bbb
I\uu a\ll b (\t + 1) = \O\ll b I\uu{a+b}\ll b
\\\\
I\uu a\ll b ( - {1\over \t}) = 
\O\ll 1\uu{-2ab} I\uu {-b}\ll{a} (\t)
\eee
\renewcommand{\fft}{I}

Despite forming a projective representation of $SL(2,\IZ / 4)$,
we can use these objects to construct a true linear representation
of $SL(2, \IZ)$ as follows.  Define $\fft\pr{}\uu a\ll 0$
as $\fft\uu a\ll 0$.  We can then define a set
of functions $\fft\pr{}\uu a\ll b$ with
no assumed periodicity on $a$ and $b$ by
defining them to represent
$PSL(2,\IZ)$ classically.
That is, we define
\bbb
\fft\pr \uu{a\pr}\ll{b\pr} (\t) \equiv
\fft\pr\uu a\ll 0 ({{A \t + B}\over{C\t + D}}) 
\eee
if
\bbb
\lsqq \bm a\pr \cr b\pr \em \rsqq
=
\lsqq \bm A & B \cr C & D \em \rsqq
 \lsqq \bm a \cr 0 \em \rsqq
\eee
The resulting objects $\fft\pr{}\uu a\ll b$
transform classically under $PSL(2,\IZ)$,
with modular transformations
acting on $a$ and $b$ as a doublet and no phases
multiplying $\fft\pr{}\uu a\ll b$.
This is automatically so, because every pair of integers is
$PSL(2,\IZ)$ equivalent to a unique element of the
form $(a,0)$, up to overall sign.  Since $PSL(2,\IZ)$
acts on $\t$, the set of all functions on the upper
half plane must consistently represent $PSL(2,\IZ)$ in
a linear way.  It does not matter
which $PSL(2,\IZ)$ transformation we use to
transform $(a\pr,0)$ to the form $(a, b)$ in order
to define the function $\fft\pr{} \uu a\ll b$; the
definition of the $\fft\pr$ will be independent
of such choices. 

It is clear from our earlier discussion that the $\fft\pr$
are equal to the corresponding
$\fft$ up to phases.  Specifically, $I\pr{}\uu a\ll b
= \lrdd \O\ll 1 \rrdd\uu {ab} I\uu a\ll b$, with
$\O\ll 1$ as defined above, $\O\ll 1 = \s \exp{{\pi i \over 4}}$.

The important
point is that the $\fft\pr{}\uu a\ll 0$ are equal to $\fft\uu a\ll 0$
so the untwisted sector partition function is the same
in terms of the $I\pr$ as in terms of the
$I$.

\heading{Interpretation}

We define the partitition function of the full theory
as $\sum\ll{a,b}  \fft\pr{}\uu a\ll b \cdot
Y\uu a\ll b$.  Now consider the meaning
of the sum for a fixed value
in $b$.  If we rewrite the $I\pr{}\uu a\ll b$ in terms
of the original traces $I\uu a\ll b$, the sum is
\bbb
Z\ll b \equiv \sum\ll a Y\uu a\ll b (\O\ll 1)\uu {a} 
I\uu a\ll b
\eee
This is equal to
\bbb
\sum\ll a \tr\ll{[w=b,~\rm{twisted~by\it g\uu b}]}
\O\ll b\uu {a} \lrdd g\ll  X\uu a g\uu a q\uu{L\ll 0}
\qb\uu{\tilde{L}\ll 0} \rrdd 
\eee
Without the phases of $\O\ll b\uu{a}$, this would be
a projection onto states with winding $b$, twisted by $g\uu b$,
satisfying $g\cdot g\ll x = 1$.  That is, states
for which the eigenvalue of $g$ was equal to the eigenvalue of
the operator $g\ll x = \exp{2\pi i R\ll {base}p\ll{base}}$,
which shifts $X\uu 7$ by $2\pi R\ll{base}$.  This would
be just as in a conventional orbifold construction.  The
effect of the phase $\O\ll b\uu{a}$ is to alter the
orbifold projection to $g \cdot g\ll x = \O\ll b\uu{-1} = \O\ll 1
\uu{-b}$
in the sector with $w=b$.  This is a direct derivation,
starting from the imposition of modular invariance,
of the anomalous momentum fractionation
which we introduced \it ad hoc \rm earlier.

We add one final note
about the phases $\O\ll b$.  The definition $I\pr{}\uu a\ll b
= \lrdd \O\ll 1 \rrdd\uu{ab} I\uu a\ll b$ and
the anomalous transformation $I\uu a\ll b(\t + 1) = \O\ll b
I\uu {a+b}\ll b (\t)$ implies $\O\ll b = \lrdd \O\ll 1 \rrdd
\uu {b}$.  This reproduces the relation we derived
in equation (\ref{genform}).  In particular, the phase
$\O\ll b = \lrdd \O\ll 1\rrdd\uu b$ represents a multiplicatively
conserved, winding-dependent modification of the 
projection in the sector $w\ll{base} = 0$.

\subsection{Spectrum of light states}

Let us use the requirement of
level matching to determine the spectrum
of states.  The simplest way
to proceed is to begin by
considering states with no momentum
or winding on either direction of
the fiber. 

Restrict attention
to states which are in the NS sector
on the right-hand side; the right moving R spectrum
can be restored at the very end by taking the OPE
with $\nsp/ \rrp$ states with no momemtum, winding
or phase under $g$.

The untwisted sector contains $\nsp/\nsp$ states
with $(n\ll 7, w\ll 7) = (a,0)$ mod $(1,8)$ and
eigenvalue $\exp{\pi i \a}$ under $g$, where $a = 0,\hh$.
There are also $\rrp/\nsp$ states of the same form
with $\a = {1\over 4}, {3\over 4}$.  It is clear,
then, that the untwisted sector has a closed OPE and
reproduces the expected boundary condition for local fields in 
the base $x\uu 7$ in terms of their $g$ transformations.

Now consider the first twisted sector, in particular the
NS/NS states.  First, note that there is a single
complex left-moving fermion which is periodic in this
sector, namely $\pst\uu 8 + i \pst\uu 9$,
so the ground state phase under $\mflw$ is $\pm i$
rather than $\pm 1$.  (If necessary this can be seen by bosonizing
the complex fermion.)  Next, we find that the level mismatch due
to Casimir energy in
this sector is $\tilde{L}\ll 0 - L\ll 0 = {1\over 4}$ in the oscillator
ground state.  However as shown in the appendix, twisted sectors
of the T-duality transformation should have the \it right- \rm moving
zero modes quantized in half-integer units.  So there is an
extra ${1\over {16}} + {1\over {16}} = {1\over 8}$
contribution to $L\ll 0$ from the zero modes $p\ll R \uu{8,9}$ which
are each equal to $\pm \hh$ in the lowest state.  

Thus the total level mismatch of the lowest state in the twisted sector
is $\tilde{L}\ll 0 - L\ll 0 = +{1\over 8}$.
This must be cancelled by a level mismatch contribution $\Delta
(\tilde{L}\ll 0 - L\ll 0) = - n\ll 7 w\ll 7$.  So we take
$n\ll 7 = {1\over 8}$ in the first twisted sector, which has $w\ll 7 = 1$.
The $g$-eigenvalue of the ground state is the same as its
$\mflw$ eigenvalue, because the fermion zero mode $\pst\ll 0\uu 8 + i
\pst\ll 0\uu 9$ gives the sole contribution to both.
So the ground state has $g =  \mfls = i$.


Now we can build the rest of the states
by taking the OPE with the untwisted sector.
In the sector with $w=1$, for instance,
we have $\nsi/\nsp$ states $(n\ll 7, w\ll 7)
= ({1\over 8} + \a,1)$ and $g$-eigenvalue $\exp{2\pi i \a}$
with $\a = {1\over 4}, {3\over 4}$.  In the $\rri/\nsp$ 
sector we get states with $(n\ll 7 , w\ll 7) = ({3\over 8} +\a,
1)$ and $g = \exp{2\pi i \a}$ where $\a$ can be $0$ or $\hh$. 

In the sector with $w\ll 7 = 2$ we get a particularly interesting
set of NS/NS states.  We can obtain them by taking the OPE of
two identical $\rpmi/\nsp$ states with $(n\ll 7, w\ll 7) = (\pm {3\over 8},
\pm 1)$
mod $(1,8)$.  The resulting sector is $\nsm/\nsp$ with $g = +1$
and $(n\ll 7, w\ll 7) = (\mp {1\over 4}, \mp 2)$
mod $(1,8)$.  This sector contains the $W\ll\pm$-bosons and 
off-diagonal
adjoint scalars
of an SU(2) gauge symmetry, of which the left-moving current
$\pp \ll - X\uu 7$ is the generically unbroken generator.  The 
states are of the form
\bbb
\psi\uu\m\ll{-\hh}\kket{p\ll 7\up L = \pm {1\over{4R\ll 7}}
  \pm {{2R\ll 7}\over{\apr}},
p\ll 7 \up R = \pm {1\over{4R\ll 7}} 
\mp {{2R\ll 7}\over{\apr}}\otimes(\rm osc.~vac.)
\otimes(\rm NS~ghost~vac.)
 }  
\eee
with $\m$ ranging from $0$ through $9$.  The components $0-6$ give
a 7D vector field and the components $6-9$ give the off-diagonal
components of the adjoint
scalar fields.
At $R\ll 7 = {{\sqrt{2\apr}}\over{4}}$ the mass-squared of the
state goes to zero and the SU(2) gauge symmetry is unbroken.  All
three components of the adjoint scalar are also massless at
this radius, with the third component being the modulus $R\ll 7$.

The physics is essentially identical to that of the ehnanced
SU(2) gauge symmetry of the chiral Scherk-Schwarz compactification
described in \cite{SHsnew}.  The only difference is
that our theory now lives in seven rather than nine noncompact
dimensions.  The reason for the similarity
is clear when one considers that the Wilson line in that
model implemented an
action of the symmetry $\mfls$, whereas the Wilson line in the
model considered in this section implements the symmetry $g$, which
squares to $\mfls$.  So it is altogether logical that the
sector twisted by $g\sqd$ in the theory we consider here 
has a massless field content resembling that of the twisted sector
in the Wilson line model of \cite{SHsnew}.

The rest of the behavior of our 7D theory parallels that
of the Wilson line of \cite{SHsnew}, and we shall not
re-derive the parallel results in detail.  Instead, we list
them briefly:

\bi
\item{The theory has sixteen supercharges in 7 dimensions.
 The massless gravitino comes from the
right-moving spin fields.  The the left-moving spin fields
give rise to fields which are antiperiodic around the $X\uu 7$ circle,
and therefore do not give rise to massless gravitini in $7$ dimensions.
However we can see that the left-moving
gravitini do become light in
the limit $R\ll 7\to\infty$.  This must be so, because when $R\ll 7$
decompactifies, the information about the Wilson line is lost, so
we must recover a theory on the moduli space of type II string theory
compactified on a $T\uu 2$, which has 32 supercharges in 8 dimensions. }
\item{The massless
field content at a generic radius is a
gravitational multiplet and a ${\cal N}= 1$ abelian
vector
multiplet.  In 7D the ${\cal N} = 1$
gravitational multiplet contains
a single scalar and three abelian vector fields.  The
vector multiplet contains three scalars and one vector field.
The four scalars come from the dilaton, the radius
of the base, and the $X\ll 7$ components of
the two right-moving abelian vectors of the fiber torus.
The three graviphotons are the
abelian vectors corresponding to the right-moving worldsheet 
currents $\pp\ll + X\uu{7,8,9}$.  The gauge field in the
vector multiplet corresponds to the current $\pp\ll - X\uu 7$
at a generic radius.}
\item{The value of $n\ll{base}$ mod ${1\over 4}$ is
determined by the periodicity of the
left-moving worldsheet supercurent $\tilde{G}$: in left-moving
NS sectors $n\ll{base}$ is equal to $0$ mod ${1\over{4}}$
and in left-moving R sectors $n\ll{base}$ is equal to ${1\over 8}$
mod ${1\over 4}$.  It follows immediately that there are no
R/NS or R/R sectors which are massless as 7-dimensional fields.
We therefore expect D-brane charges in 7D to be conserved 
at most modulo finite integers.  This is a similar
situation to that in the chiral Scherk-Schwarz compactification
of (\cite{SHsnew}).}
\item{The theory has a \it self-\rm T-duality which 
takes $R\ll 7$ to $\apr / 8 R\ll 7$.  Just as in (\cite{SHsnew}),
this
is a duality between type IIA and itself, or type IIB and itself,
rather than exchanging the two type II theories.}
\item{At the self-dual radius the theory has an enhanced SU(2) gauge
symmetry with a complete massless vector multiplet
of SU(2).}
\ei


\section{Conclusions}\label{conclusions}

In this paper we have demonstrated the consistency of
Wilson line backgrounds for a range of stringy
symmetries.  The Wilson lines implement monodromies by
symmetries $g$ as parallel transport around a base.
The symmetries $g$ are symmetries of a compact internal
'fiber' string theory (in all cases we consider, a torus or 
some type of current algebra) at a point in moduli space where
$g$ is unbroken.

We contrast the class of models we consider to the case where one
considers a Wilson line for a symmetry $g$ which is spontaneously
broken in the fiber theory.  (See for
example \cite{STW}, 

In such a case, there must always
be gradients of the fiber moduli over the base $S\uu 1$, because
the moduli ${\cal M}$ must interpolate between ${\cal M}\ll 0$
and ${\cal M}\pr \equiv g\cdot {\cal M}\ll 0$ as one goes around
the base.  This means the theory must always have positive energy, which
cannot be cancelled without positive curvature, orientifolds, or
some other source of negative energy.  The case
where the gradient energies are cancelled with a positively curved
two-dimensional base (with singular points) was considered in \cite{HMW}
and the case where the energy is cancelled by orientifolds was considered
in \cite{STW}.  However in neither case is the theory described by
a free worldsheet CFT with an arbitrarily weak dilaton.
\footnote{In special points of the moduli spaces in \cite{HMW}, the
moduli of the fiber are fixed under the monodromy group, and
in those cases there appears to be a free worldsheet CFT describing
string propagation.}

In
our models, by contrast, the full moduli space is described by free,
solvable CFT.  This allows us to go beyond the large-base limit,
or equivalently the adiabatic approximation for the dynamics of
the fiber.  Indeed, when the base to be smaller than string scale,
interesting new phenomena appear, including
enhanced gauge symmetry and self-T-duality.

To demonstrate the consistency of our theories, we calculated
several partition functions, checking their modular invariance.
We would like to emphasize particularly that the stringy Wilson line
theories make sense even for symmetries $g$ which would lead to 
anomalous, inconsistent theories if one were to attempt simply
to orbifold by $g$.

We draw special attention to a distinction between
the 'tame' and 'wild' classes of Wilson line.  The distinction
can be expressed in terms of the denominator of $n\ll {base}\equiv p\ll
{base} R\ll {base}$ and the order $N$ of the symmetry $g$.  For tame
Wilson lines, the momentum in the base is always an integer multiple
of ${1\over {N R\ll{ base}}}$, in the twisted (winding) as well as untwisted sectors.
Consequently these theories can be understood in the obvious way, as 
orbifolds of the $N$-fold cover of the base by a $2\pi R\ll{base}$ shift,
combined with the action of $g$. For the wild case, the 
momentum $n\ll 9$ in the winding sectors can have a denominator
as large as $N\sqd$.  We have seen that in order to interpret
these wild Wilson lines as quotionts of an $N$-fold cover, the
orbifold action must contain additional phases which do not factorize
between base and fiber.

The guiding principle in our constructions was the 'CCC principle':
consistent boundary conditions for consistent string theories lead
to consistent backgrounds.  Though almost a tautology, the CCC principle
is an extremely useful analytical tool when one wishes
to find a description whose spacetime interpretation (in the
effective theory after reduction on the fiber) is straightforward
but whose worldsheet description may nonetheless be quite involved.
The use of the CCC principle has allowed us to derive a detailed
worldsheet descriptions of a number of new theories.  As in
the orbifold constructions of \cite{vafa1, vafa2}, level
matching and closure of the OPE turns out to be a
sufficient condition for modular invariance in all
cases.

Most interestingly,
our type II model
preserves sixteen supercharges in seven dimensions and
its moduli space does not appear to be equivalent to the moduli space of 
any previously known compactification. In particular, it does not seem
to be (perturbatively or non-perturbatively) connected to any of the
moduli spaces on the list of, {\it eg}, ref.\ \cite{triples}. The simplest
way to see this is to look at the rank of the gauge group.
 

The existence of highly supersymmetric, solvable nongeometric backgrounds
is encouraging.  It may be possible to fiber these examples further over 
a base to obtain lower-dimensional models, and/or to break some more
SUSY by adding branes, fluxes and orientifolds to obtain interesting 
and controllable ${\cal N} = 1$ compactifications.  As has been
frequently pointed out \cite{HMW,hellerman,STW, LSW} nongeometric 
compactifications inevitably dominate the string landscape, even 
in ensembles preserving some low energy supersymmetry.  This frontier
is now open for exploration.

\begin{acknowledgments}
We would like to thank Jacques Distler, Alex Flournoy, Albion
Lawrence, Michael 
Schulz, Nathan Seiberg, Washington Taylor, Brook Williams, and Brian 
Wecht for valuable discussions and comments. This work was supported 
in part by the DOE under grant number DE-FG02-90ER40542.
S.H. would also like to thank the Perimeter
institute and the Harvard theory group for
hospitality while this work was being completed.  S.H. is the
D. E. Shaw \& Co., L. P. Member at the Institute for
Advanced Study.
\end{acknowledgments}

\vfill\eject

\appendix{}

\section{Cocycles and phases in the T-duality transformation}
\label{cocycles}

In this section of the Appendix we will
develop a few technical components used in our description
of Wilson lines for T-duality transformations on
tori in the bosonic string and the superstring.
First we will review the theory of the
fiber circle $S\uu 1$
at the self-dual radius in the bosonic string.

\subsection{Local operators and cocycles in toroidal compactification}

Consider the internal CFT ${\cal I}$, where ${\cal I}$ is
a fiber circle $S\uu 1$, described by the $c=1$ CFT
of a circle at the self-dual radius $\sqrt{\apr}$.  
Each state is characterized by a momentum $\nf$ and
a winding number $\wf$, as well as a set of left-moving
and right-moving occupation numbers
$\Nft\uu i$ and $\Nf\uu i$.
(When there is no danger of confusion
we will drop the subscript and write $n,w, \tilde{N}\uu{i}$
and $N\uu i$.)  It is often convenient to recombine the
quantum numbers $n,w$ into the chiral quantum numbers
$p\ll{R,L}$, which are related by
\bbb
p\ll L \equiv n - w
\llsk\llsk\llsk
p\ll R \equiv n + w
\\\\
n = \hh(p\ll L + p\ll R)
\llsk\llsk\llsk
w = \hh (- p\ll L + p\ll R)
\eee
The conformal weights of the ground state of a
sector with
given momentum and winding are
\bbb
\htt\ll{(n,w)} = {1\over 4} (n - w)\sqd = {1\over 4}
p\ll L\sqd
\llsk\llsk\llsk
h\ll{(n,w)} = {1\over 4} (n + w)\sqd =
{1\over 4}
 p\ll R\sqd
\eee
T-duality (for a review see \cite{tdual}) is an unbroken symmetry of 
strings on a circle of radius $\sqrt{\apr}$ which inverts the left-moving 
part $X\uu L$ of the coordinate of the circle:
\bbb
X\uu L\to - X\uu L \llsk\llsk\llsk X\uu R \to + X\uu R.
\eee
This operation 
switches the quantum numbers $n$ and $w$, as well as acting with
a sign $(-1)\uu {\Nt}\equiv (-1)\uu {\sum\ll i \tilde{N}}$
depending on the number of left-moving oscillators
excited in a given state.
The symmetry we have just defined, which we shall call
$T$, is cyclic of order 2 and is a symmetry
of the conformal field theory.

We can decompose our space of states into two sectors, which we
call ${\cal U}
\uu{\pm}$, where the sign $\pm$ labels the
eigenvalue of a sector under $T$.  The ${\cal U}$ denotes the fact
that all states in ${\cal I}$ have
untwisted, periodic boundary conditions for 
$X\uu L$; their spin ${\htt}
- h$ is equal to zero modulo the integers.

The discrete symmetry $X\uu L \to - X\uu L$
corresponds to an action on Hilbert space
\bbb
\kket{n,w,\tilde{N}\uu i, N\uu i}
\to \Omega\ll{n,w} ~(-1)\uu{\sum \tilde{N}\ll i}~
\kket{w,n,\tilde{N}\uu i, N\uu i}
\eee
which switches $n$ and $w$, and anticommutes with every
left-moving oscillator $\tilde{\a}$.  The 
only freedom is a possible phase $\Omega\ll{n,w}$ which
depends on $n$ and $w$ and for consistency must
satisfy $\Omega\ll{n,w} \Omega\ll{w,n} = 1$.

\subsection{Conserved and non-conserved T-duality operations}

The simplest choice of phase would be $\Omega\ll{w,n} = 1$, and
the corresponding naive $T$-duality
operation on the single-string Hilbert space
is a symmetry of the spectrum for all $w,n,\tilde{N}\ll i$.
However it is \it not \rm
a symmetry of the CFT as a whole.  In particular,
this symmetry is not preserved by the operator product
expansion.  Therefore $\Omega\ll{w,n} = 1$ does
not correspond to a symmetry which is conserved in
string scattering interactions.

Let us see why the naive T-duality symmetry, which we will
call $T\ll 0$, is not preserved by the structure of the OPE.
The origin of the 
nonconservation of $T\ll 0$ is in the cocycle
factor in the definition of vertex operators
carrying momentum and winding.  (For an introduction, see
for instance(\cite{joe})).
The naive expressions
\bbb
V\up 0\ll{(n,w)} = \exp{i (n-w) X\ll L + i (n +w) X\ll R} 
= \exp{i p\ll L X\ll L + i p\ll R X\ll R}
\eee
are not mutually local and are therefore cannot
be good conformal fields corresponding to states via the
state-operator correspondence.

To see that they are not local, we review the
argument in \cite{joe}.  The operator $X\ll L$ is not
local with respect to itself, and similarly for $X\ll R$.  The
OPE of a chiral scalar with itself has a logarithmic
singularity, which we can give a definite value by
letting the branch cut run along the negative imaginary direction
from the location of each operator.
Then
\begin{equation*}
V\up 0\ll{(n,w)}(\s\ll 1 + i\e) V\up 0\ll{(n\pr, w\pr)}(0)
=
\exp{\pi i (n w\pr + w n\pr)}
V\up 0\ll{(n,w)}(\s\ll 1 - i\e) V\up 0\ll{(n\pr, w\pr)}(0)
\end{equation*}
for $\s\ll 1,\e \in \IR$ and $\e\to 0\uu +$.
Wick rotating, and defining
the Lorentzian timelike direction as $i \s\ll 2 = i {\rm Im} z$,
the discontinuity becomes a failure of 
the spacelike separated operators $V\up 0\ll{(n,w)}(\s\ll 1,t)$
and $V\up 0\ll{(n\pr, w\pr)}(\s\ll 1,t)$ to commute:
\begin{equation*}
V\up 0\ll{(n,w)}(\s\ll 1,t)~V\up 0\ll{(n\pr, w\pr)}(\s\ll 1,t)
=
\exp{\pi i (n w\pr + w n\pr)}
V\up 0\ll{(n\pr, w\pr)}(\s\ll 1,t)~V\up 0\ll{(n,w)}(\s\ll 1,t)
\end{equation*}

To construct the appropriate local operators, multiply the
naive expression for
the vertex operator by the nonlocal expression $\Ch\ll {(n,w)}$:
\bbb
V\ll{(n,w)} = \hat{C}\ll{(n,w)}\cdot V\up 0\ll{(n,w)}
\eee
where $\Ch\ll{(n,w)}$ is a nonlocal expression built
from the zero modes of the field $\xf$:
\bbb
\Ch\ll{(n,w)} \equiv \exp{\pi i w \hat{n} - {{\pi i}\over 2} nw} 
\eee
and $V\up 0\ll{(n,w)}$ is the naive expression given
above for the
vertex operator.

Here, the symbol $\hat{n}$ denotes the \it operator \rm
which acts on the Hilbert space as an infinitesimal
translation, and $n,w$ are the \it c-numbers \rm
which label the vertex operator $V\ll{(n,w)}$.
Concretely, $\Ch\ll{(n,w)}$ is equal to $i\uu{nw}$ if $w$ is even
and $i\uu{nw}\cdot (-1)\uu{\hat{n}}$ if $w$ is odd.\footnote{The
phase $i\uu{-nw}$ does not appear in the cocycle as defined
in \cite{joe}.  We include
it so that the full vertex operator including the cocycle
will obey the standard hermiticity condition $V\dag\ll{(n,w)}
= V\ll{(-n,-w)}$.  Our choice also makes
the T-duality properties
of our vertex operators simpler than those with the
phase convention in \cite{joe}.}

This nonlocal operator $\Ch\ll{(n,w)}$ cancels the nonlocality
in the OPE of the naive vertex operators $V \ll{(n,w)}$
with each other.  There is some arbitrariness
in the choice of $\hat{C}\ll{(n,w)}$.  Its only
necessary properties are that it be made only
from the operators $\hat{p}\ll{L,R}$ and that it satisfy
\bbb
\hat{C}\ll{(n,w)}~V\up 0 \ll{(n\pr,w\pr)}
=
\g\ll{(n,w|n\pr,w\pr)}V\up 0 \ll{(n\pr,w\pr)}\hat{C}\ll{(n,w)}
\\\\
{\rm with}~ \g\ll{(n,w|n\pr,w\pr)} ~{\rm a~phase~satisfying}
\\\\
\g\ll{(n,w|n\pr, w\pr)} \g\st\ll{(n\pr,w\pr|n,w)} =
\exp{\pi i (n\pr w - w\pr n )}
\een{cocycleproperty}
The redefined vertex operators are good,
mutually local conformal fields.
With the effect of the cocycle taken into account, the
OPE of two vertex operators is local:
\bbb
V\ll{(n,w)} \cdot V\ll{(n\pr, w\pr)}
\sim  \exp{{{\pi i}\over 2} (w\pr n - n\pr w)}
(\Delta \zb)\uu{\hh 
p\ll L p\pr\ll L} (\Delta z)\uu{\hh p\ll R p\pr\ll R}.
\eee
That is, if we deifine fractional powers of $\Delta z$
with the branch cuts running from $\Delta z = 0$ to $\Delta z
= - i \infty$, then there is no branch cut in the
overall OPE, as long as $n$ and $w$ are integers.

Now it is easy to check explicitly
that the naive $\IZ\ll 2$ T-duality
operation $T\ll 0$ is \it not \rm respected by the OPE, and
therefore is not a symmetry of string interactions.  Suppose
we prepare states which are even under $T\ll 0$: let
\bbb
\tilde{V}\uu +\ll{(n,w)}
\equiv {1\over{\sqrt{2}}} \lrdd 
\tilde{V}\ll{(n,w)} + V\ll{(w,n)} \rrdd.
\eee
The vertex operators $\tilde{V}
\uu +\ll{(n,w)}$ are even under $T\ll 0$,
by construction.  If the symmetry operation
$T\ll 0$ were respected by the
OPE, then the OPE of two $\tilde{V}\uu +$ operators would have to 
contain only vertex operators which are even under $T\ll 0$.
But we can see that this is not the case.  Using the OPE
we just derived, we have
\bbb
\tilde{V}\uu +\ll{(n,w)} (z\ll 1, \zb\ll 1) \tilde{V}
\uu + \ll{(n\pr, w\pr)}
(z\ll 2, \zb\ll 2) 
\\\\
\sim \hh ~\zb\ll{12}
\uu {\hh k\ll L k\pr\ll L} z\ll{12}\uu {\hh k\ll R k\pr\ll R}
\lrdd i\uu{w\pr n - n\pr w} V\ll{(n + n\pr, w + w\pr)}
+ i\uu{n\pr w - w\pr n} V\ll{(w + w\pr,n + n\pr)} \rrdd
\\\\
+ \hh ~\zb\ll{12}
\uu {- \hh k\ll L k\pr\ll L} z\ll{12}\uu {\hh k\ll R k\pr\ll R}
\lrdd
  i\uu{n\pr n - w\pr w} V\ll{(n + w\pr, w + n\pr)}
+ i\uu{w\pr w - n\pr n} V\ll{(w + n\pr , n + w\pr)}
\rrdd.
\eee
The right hand side of the OPE contains the terms
\bbb
V\ll{(n + n\pr, w + w\pr)} + 
(-1)
\uu{n\pr w - w\pr n}  V\ll{(w + w\pr,n + n\pr)}
\eee  
and
\bbb
V\ll{(n + w\pr, w + n\pr)}
+ (-1)\uu{w\pr w - n\pr n} V\ll{(w + n\pr , n + w\pr)},
\eee
which is even under $T\ll 0$ only if $n\pr w - w\pr n$ and
$n\pr n - w\pr w$ are both even, which is not the case
in general.

Let us instead organize states according to their eigenvalues
under the operation $T\equiv T\ll 0 \cdot (-1)\uu{\nh\wh}
= (-1)\uu{\nh\wh}\cdot T\ll 0$.  
The operator $T$ is also a symmetry of the spectrum, but
unlike $T\ll 0$ we will now see it is also
preserved by the OPE.

Define 
\bbb
V\uu \pm\ll{(n,w)}
 \equiv 
{1\over{\sqrt{2}}}  ( V\ll{(n,w)} \pm T\cdot V\ll{(w,n)} )
\eee
The states $V\uu \pm\ll{(n,w)}$
have definite conformal weight $(\tilde{h}\ll{(n,w)}, h\ll{(n,w)})$
and definite eigenvalue $\pm$ under the redefined
T-duality operation $T$.  Note that $V\ll{(n,w)}$ is
proportional but necessarily equal to $V\uu\pm\ll{(n,w)}$,
and that some of the $V\uu\pm\ll{(n,w)}$ may vanish, when $w = n$.

Using our definition of $V\uu \pm\ll{(n,w)}$ and our OPE for the 
$V\ll{(n,w)}$, letting $\O,\O\pr$
take values in $\{\pm 1\}$
we find that the OPE of $V\uu\O\ll{(n,w)}$
with $V\uu {\O\pr}\ll{(n\pr,w\pr)}$ contains the
primaries
\bbb
\a\ll 1 V\ll{(n + n\pr, w + w\pr)} + \a\ll 2 V\ll{(w + w\pr,
n + n\pr)}
\eee
and
\bbb
\b\ll 1 V\ll{(n + w\pr, w + n\pr)} + \b\ll 2 V\ll{(w + n\pr,
n + w\pr)}
\eee
and their descendents,
where
\bbb
\a\ll 1 \equiv 
\hh
~i\uu{n w\pr - w n\pr}
\\\\
\a\ll 2 = \hh ~ 
i \uu{w n\pr - n w \pr} (-1)\uu{ n w + n\pr w\pr}
\cdot \O \cdot \O\pr 
\\\\
\b\ll 1 \equiv \hh ~ \O\pr ~ (-1)\uu{n\pr + n\pr w\pr}~
i\uu{n n\pr - w w\pr}
\\\\
\b\ll 2 \equiv \hh ~ \O  (-1)\uu{ n w}~
i\uu{w w\pr - n n\pr} 
\eee

The overall normalization of the vertex operators on the
RHS of the OPE is not of interest to us; therefore we only really
care about the relative phase between $\a\ll 1$ and $\a\ll 2$,
and similarly for $\b\ll{1,2}$.  We have
\bbb
\a\ll 2 = (-1)\uu{ w n\pr - n w\pr + n n\pr + w w\pr}
 \cdot \O \cdot \O\pr 
\cdot \a\ll 1
\\\\
= (-1)\uu {(n + n\pr)(w + w\pr)} (\O\O\pr) \a\ll 1
\eee
and
\bbb
\b\ll 2 = (-1)\uu{ n\pr w\pr + 
n w + w w\pr - n n\pr} ~ \O \O\pr  ~ \b\ll 1
\\\\
= (-1)\uu{(n + w\pr)(w + n\pr)} (\O\O\pr)
~\b\ll 2~
\eee

So the combinations which occur are precisely
\bbb
V\ll {(n + n\pr, w + w\pr)} + (\O\O\pr)
(-1)\uu{(n + n\pr)(w + w\pr)} V\ll{(w + w\pr,
n + n\pr)} = \sqrt{2} ~V\uu{\O\O\pr}\ll{(n + n\pr, w + w\pr)}
\eee
and
\bbb
V\ll {(n + w\pr , w + n\pr)} + (\O\O\pr)
(-1)\uu{(n + w\pr)(w + n\pr)} V\ll{(w + n\pr,
n + w\pr)} = \sqrt{2} ~V\uu{\O\O\pr}\ll{(n + w\pr, w + n\pr)}
\eee

The important point is that the OPE of two primaries with
$T$-eigenvalues $\O$ and $\O\pr$ contains only
primaries with $T$-eigenvalue $\O\O\pr$; in other words,
the $\IZ\ll 2$ symmetry group generated by $T$
is preserved by the OPE.

\subsection{Twisted sectors of the conserved T-duality}

In order
to define the Wilson line
CFT, it is
necessary to
extend the action of $T$ from
the theory of the ${\cal I}$ circle by itself, to
states with twisted boundary conditions on $X\uu L$ -- that is,
boundary conditions in which the field $X$ returns to
itself up to a T-duality transformation.
We need to extend $T$ in such a way that the OPE
including twisted states will preserve $T$.  Our ability
to extend $T$ as an operation of order two is
equivalent to the property that the Wilson line CFT in
the bosonic string is tame.
   
In a twisted state $\ct$
the left-moving oscillators $\at$
are half-integrally moded with frequencies
$a + \hh, a\in \IZ, a \geq 0$.  The left-moving
zero mode $x\uu L\ll 0$ is not present in this sector, so
there is no $p\ll L$.  Consequently $n = w$ in the twisted
sector.

The twisted
sectors have a Casimir momentum
$\tilde{L}\ll 0 - L\ll 0 \in \IZ + {1\over{16}}$,
which we cancel by an appropriate quantization of
the value of $p\ll R = n = w$ in the twisted sector,
as discussed in section \ref{bosonic}:
\bbb
p\ll R = \pm\hh \llsk\llsk\llsk \lrdd {\rm mod~} 1\rrdd
\eee

The resulting twisted states satisfy level matching mod $\hh$ in
the bosonic string, without the addition of
a base direction $X\ll{base}$.  The level mismatch
of $\hh$ can be made up by acting with left-moving
oscillators, which are
half-integrally moded in the twisted sector.

In
the next section of the Appendix we derive this quantization rule
from an explicit modular transformation
of the untwisted sector partition function with
an insertion of $T$.  For now, we simply
assume this quantization rule for
purposes of checking the
consistency of operator product expansions involving
two untwisted states and a twisted state.

\heading{An expression for the cocycle in the
twisted sector}

Our
operator expression
for untwisted vertex operators cannot
be applied as an operator
which acts on twisted states,
since it contains an operator $\hat{p}\ll L$
which is not defined in the twisted sector.
Therefore we need to find a definition 
of the untwisted operator $V\ll{(n,w)}$ which
renders any two vertex operators local with
respect to one another, yet can be defined 
as operators which act on twisted states.  This is necessary
if the OPE of a twisted and an untwisted state is to
be given an unambiguous form.

Our expression for the cocycle in the twisted sector is
something which
can involve the labels $n,w \simeq k\ll L, k\ll R$ of the
untwisted state, but
can involve only the \it Hilbert space operator \rm $\hat{p}\ll R$
and not the operator $\hat{p}\ll L$ because $\hat{p}\ll L$ does not
exist in the twisted sector.

We find that
the candidate cocycle
\bbb
C\up T\ll{(n,w)} \equiv
\exp{- {{\pi i}\over 2} k\ll L \hat{p}\ll R
+ {{\pi i }\over 4} k\ll L k\ll R}
\eee
satisfies the necessary condition (\ref{cocycleproperty}).
This means that the vertex operators
\bbb
V\ll{(n,w)} \equiv C\up T\ll{(n,w)}~V\up 0\ll{(n,w)}
\eee
are mutually local with respect to one another in
the presence of a twist field.

We also need an expression for $T$ itself
in the twisted sector which is
consistent with multiplication by untwisted 
local operators.  We would expect the consistent
definition to be such that $T = 1$ for
the twisted states which are level matched
mod 1, and $T = -1$ for states with
level mismatch equal to $hh$ mod 1.  A
candidate $T$ operator
in the twisted sector is
\bbb
T\up T = (-1)\uu{\lrdd \sum\ll{m = 0}\uu\infty \tilde{N}\ll 
{m + \hh} \rrdd} \exp{{{\pi i }\over 2}(\hat{p}\ll R\sqd 
- {1\over 4}) }
\eee
As in the untwisted sector, the exponent is
quadratic in the zero modes of $X$.

It is not clear \it a priori \rm that the definition of
$T$ in the twisted sectors is consistent with
the definition in the
untwisted sectors.  We can establish this
by showing that the $T$-even and $T$-odd twisted
states are separately modules over $T$-even untwisted
vertex operators, defined with the use of our
expression for the cocycle.

We can show this by explicit calculation.
Let us
summarize the action of $T$-even
untwisted vertex operators $U\uu
{( +)}
\ll{(p\ll R \equiv 1)}$
with $p\ll R$ equal to $1$ (mod 4) on twisted states.  We have:
\bbb
U\uu
{( +)}
\ll{(p\ll R \equiv 1)}
~\lrdd \bm
  \kket{p\ll R \equiv -\hh ~(\mod 4) ,~N\ll L 
\equiv a ~(\mod 2)}
\cr
\kket{p\ll R \equiv +\hh ~(\mod 4) ,~N\ll L 
\equiv a~ (\mod 2)} 
\cr
\kket{p\ll R \equiv +{3\over 2} ~(\mod 4) ,~N\ll L 
\equiv a ~(\mod 2)} 
\cr
\kket{p\ll R \equiv +{5\over 2}~ (\mod 4) ,~N\ll L 
\equiv a~ (\mod 2)} 
\em
\rrdd
\\\\
 = \lrdd \bm  \kket{p\ll R \equiv +\hh~ (\mod 4) ,~N\ll L 
\equiv a ~(\mod 2)} 
\cr
\kket{p\ll R \equiv +{3\over 2} ~(\mod 4) ,~N\ll L 
\equiv  a+1~ (\mod 2)}
\cr
\kket{p\ll R \equiv +{5\over 2} ~(\mod 4) ,~N\ll L 
\equiv  a ~(\mod 2)} 
\cr
\kket{p\ll R \equiv -{1\over 2}~ (\mod 4) ,~N\ll L 
\equiv  a+1~ (\mod 2)} 
\em
\rrdd
\eee 
The other $T$-even untwisted vertex operators act as
\bbb
U\uu
{( +)}
\ll{(p\ll R \equiv 2)}
~\lrdd \bm  \kket{p\ll R \equiv -\hh ~(\mod 4) ,~N\ll L 
\equiv a ~(\mod 2)} 
\cr
\kket{p\ll R \equiv +\hh ~(\mod 4) ,~N\ll L 
\equiv a~ (\mod 2)} 
\cr
\kket{p\ll R \equiv +{3\over 2} ~(\mod 4) ,~N\ll L 
\equiv a ~(\mod 2)} 
\cr
\kket{p\ll R \equiv +{5\over 2}~ (\mod 4) ,~N\ll L 
\equiv a~ (\mod 2)} 
\em
\rrdd
\\\\
 = \lrdd \bm  \kket{p\ll R \equiv +{3\over 2}
~ (\mod 4) ,~N\ll L 
\equiv a+1 ~(\mod 2)} 
\cr
\kket{p\ll R \equiv +{5\over 2} ~(\mod 4) ,~N\ll L 
\equiv  a+1~ (\mod 2)} 
\cr
\kket{p\ll R \equiv -{1\over 2} ~(\mod 4) ,~N\ll L 
\equiv  a+1 ~(\mod 2)} 
\cr
\kket{p\ll R \equiv +{1\over 2}~ (\mod 4) ,~N\ll L 
\equiv  a+1~ (\mod 2)}  \em \rrdd
\\\\
\eee
\bbb
U\uu
{( +)}
\ll{(p\ll R \equiv 3)}
~\lrdd \bm  \kket{p\ll R \equiv -\hh ~(\mod 4) ,~N\ll L 
\equiv a ~(\mod 2)} 
\cr
\kket{p\ll R \equiv +\hh ~(\mod 4) ,~N\ll L 
\equiv a~ (\mod 2)} 
\cr
\kket{p\ll R \equiv +{3\over 2} ~(\mod 4) ,~N\ll L 
\equiv a ~(\mod 2)} 
\cr
\kket{p\ll R \equiv +{5\over 2}~ (\mod 4) ,~N\ll L 
\equiv a~ (\mod 2)} 
\em
\rrdd
\\\\
 = \lrdd \bm  \kket{p\ll R \equiv +{5\over 2}
~ (\mod 4) ,~N\ll L 
\equiv a+1 ~(\mod 2)} 
\cr
\kket{p\ll R \equiv -{1\over 2} ~(\mod 4) ,~N\ll L 
\equiv  a~ (\mod 2)} 
\cr
\kket{p\ll R \equiv +{1\over 2} ~(\mod 4) ,~N\ll L 
\equiv  a+1 ~(\mod 2)} 
\cr
\kket{p\ll R \equiv +{3\over 2}~ (\mod 4) ,~N\ll L 
\equiv  a~ (\mod 2)}  \em \rrdd
\\\\
\eee
\bbb
U\uu
{( +)}
\ll{(p\ll R \equiv 0)}
~\lrdd \bm  \kket{p\ll R \equiv -\hh ~(\mod 4) ,~N\ll L 
\equiv a ~(\mod 2)} 
\cr
\kket{p\ll R \equiv +\hh ~(\mod 4) ,~N\ll L 
\equiv a~ (\mod 2)} 
\cr
\kket{p\ll R \equiv +{3\over 2} ~(\mod 4) ,~N\ll L 
\equiv a ~(\mod 2)} 
\cr
\kket{p\ll R \equiv +{5\over 2}~ (\mod 4) ,~N\ll L 
\equiv a~ (\mod 2)} 
\em
\rrdd
\\\\
 = 
\lrdd \bm  \kket{p\ll R \equiv -\hh ~(\mod 4) ,~N\ll L 
\equiv a ~(\mod 2)} 
\cr
\kket{p\ll R \equiv +\hh ~(\mod 4) ,~N\ll L 
\equiv a~ (\mod 2)} 
\cr
\kket{p\ll R \equiv +{3\over 2} ~(\mod 4) ,~N\ll L 
\equiv a ~(\mod 2)} 
\cr
\kket{p\ll R \equiv +{5\over 2}~ (\mod 4) ,~N\ll L 
\equiv a~ (\mod 2)} 
\em \rrdd
\eee

In particular, there exist two complementary sets of twisted states
which are closed under the action of all $T$-even
untwisted vertex operators $V\up +$; namely the set of
level matched states
\bbb
\left \{~\kket {p\ll R \equiv \pm\hh
(\mod 4), N\ll L = {\rm even} }~\right \} \bigoplus
\left \{ ~\kket {p\ll R \equiv \pm {3\over 2}
(\mod 4), N\ll L = {\rm odd} }~ \right \}
\eee
and the set of un-level-matched states
\bbb
\left \{~\kket {p\ll R \equiv \pm\hh
(\mod 4), N\ll L = {\rm odd} }~\right \} \bigoplus
\left \{ ~\kket {p\ll R \equiv \pm {3\over 2} 
(\mod 4), N\ll L = {\rm even} }~ \right \}
\eee

Therefore we can define the orbifold projection in the twisted sector
as the restriction to the first set.  Level-matching follows immediately,
as does closure of the untwisted-twisted vertex operator algebra.
It remains to be shown is that two allowed twisted vertex operators
close on an allowed untwisted vertex operators.
To see this, we note the existence of
an quantum number, namely $p\ll L$ mod 1, which
commutes with T-duality.  In order for the OPE to be
potentially consistent, the value of $p\ll L$ mod 1 in
the twisted sectors must be equal to $0$ mod $\hh$.
As a result, $\exp{2\pi i p\ll L}$ never changes under
the operation $p\ll L\to - p\ll L$, meaning that the
operator $ \exp{2\pi i p\ll L}$, which measures 
$p\ll L$ mod 1 commutes with T-duality and therefore can
be assigned a definite value in any twisted sector.
This, despite the fact that $p\ll L$ is nondynamical
in the twisted sectors and does not contribute to the
energy or the Virasoro generators.  To be consistent,
then, the twisted sectors are paired with $p\ll R$
values in such a way that $p\ll R - p\ll L$ is always
even, in the twisted as well as untwisted sectors.
This guarantees the closure consistency of the twisted-twisted
OPE onto untwisted sectors.

In the bosonic string, the orbifold
by T-duality is the
same as the unorbifolded circle theory at twice the self-dual 
radius -- equivalently, the interval theory at the special
radius where it is equivalent to a circle theory.
To see this, observe that
the spectrum is left-right symmetric
at low levels -- note for instance the currents are
$U(1)\ll L \times U(1)\ll R$ and nothing more.  In fact
the T-duality transformation, with the cocycle
taken carefully into account, is a \it nonchiral \rm
symmetry, conjugate to a diagonal $\IZ\ll 2$ of
the $SU(2)\ll L \times SU(2)\ll R$ symmetry.

When embedded as a symmetry of the superstring, however, the
T-duality operation we describe here is chiral
and therefore \it not \rm conjugate to a
diagonal $\IZ\ll 2$.

\section{Fiber partition functions for T-duality Wilson lines}

Here we work out the partition functions in
twisted and untwisted sectors with
insertions of T-duality transformations for worldsheet
bosons and fermions.

\subsection{Bosonic partition functions}

In the untwisted sector, our projection again correlates the phase of
$g$ with the momentum $n\ll 7$ mod 1 in the obvious way. 
Let us construct the partition function, enforcing modular
invariance by hand.  Here $Y\uu a\ll b$ is the path integral sector
with $X\uu 7$ winding $b$ times around the spacelike and $a$ times
around the timelike cycle of the torus.
Meanwhile we would like $I\uu p\ll q$ to represent something like
the path integral with a '$g\uu p$ cut' along the timelike
cycle and a '$g\uu q$ cut' along the spacelike cycle.
Unfortunately, $g$ and its powers are really 'quantum' rather than
classical operations, so it is far from clear what it should mean to
define a path integral with cuts by $g\uu p$ and $g\uu q$ along cycles.
However objects such as $I\uu p\ll q$ have been sufficiently useful
in our study of Wilson line CFT for free fermion theories that we are
tempted to try to define an analog here.

Here is how we do it.  We think we know how the multiplicatively conserved
$g$ charge is defined, at least in the untwisted sector.  This gives
a definition to $I\uu p\ll 0$.  Now, in order to modular
transform, we want to define $I\uu 0\ll p$ with a coefficient
such that it is a sum of $q\uu h \qb\uu \htt$ with positive integer
coefficients.  {} From there we can continue.

So let us do this, first, for the simple T-duality Wilson line in
the bosonic string.  So how do we define $I\uu p\ll 0$?
Well let's begin by factorizing $I\uu 0\ll 0$ which is just
the partition function for the circle at the self-T-dual radius.
\bbb
I\uu 0\ll 0 = b~B\uu 0\ll 0~z\uu 0\ll 0
\eee
where $b$ is the partition function for the right-moving 
degrees of freedom:
\bbb
b\equiv q\uu{-{1\over{24}}} ~\prod \ll{n = 1}\uu\infty (1 - q\uu n)\uu{-1}
\eee
and $B\uu 0\ll 0$ is the corresponding partition function for
left-moving degrees of freedom
\bbb
B\uu 0\ll 0 \equiv \qb\uu{-{1\over{24}}} ~\prod \ll{n = 1}\uu\infty 
(1 - \qb\uu n)\uu{-1}
\eee
and $z\uu 0\ll 0$ is the zero mode partition function:
\bbb
z\uu 0\ll 0 = \sum\ll{n,w} q\uu{\ff(n+w)\sqd} \qb\uu{\ff(n - w)\sqd}
\eee

Now let us define the functions with 'T-duality cuts' along the timelike
direction.  This just means inserting a $-1$ into the partition functions
for each T-duality odd state.  As discussed earlier, the one non-obvious
aspect to this is that there is a factor of $(-1)\uu{nw}$ in the 
definition of the correct T-duality in the zero mode partition function.
So 
\bbb
z\uu 1\ll 0  = \sum\ll n (-1)\uu{n\sqd}
q\uu{n\sqd} 
\eee
and of course
\bbb
B\uu 1\ll 0 = \qb\uu{-{1\over{24}}} ~\prod \ll{n = 1}\uu\infty 
(1 + \qb\uu n)\uu{-1}
\eee

Now, we will insert a phase into the definition of $z\ll 1\uu 0$
and of $B\ll 1\uu 0$ so that they are sums of powers of $q,\qb$ with
positive integer coefficients.  We can perform the transformations
by noting that these sums are equal to certain things, and using
their known modular properties.
For instance,
\bbb
b(\t) = \eta(\t)\uu{-1}
\\\\
z\uu 0\ll 0 (\t) = \th\ll{00}(0,2\tb)\th\ll{00}(0,2\t)
+ \th\ll{10}(0,2\tb)\th\ll{10}(0,2\t)
\\\\
z\uu 1\ll 0 (\t) = \th\ll{01} (0,2\t)
\\\\
B\uu 0\ll 0 (\t) = \eta(\tb)\uu{-1}
\\\\
B\uu 1 \ll 0 (\t) = \lrdd {{2\eta(\tb)}\over{\th\ll{10}(0,\tb)}} \rrdd \uu\hh
\eee
and then define
\bbb
I\uu p\ll q \equiv b~z\uu p\ll q~B\uu p\ll q
\eee
First let us verify the $S$ and $T$ transformations of $I\uu 0\ll 0$.
We have
\bbb
z\uu 0\ll 0 \to \hh ~|-i\t|
\lrdd
\th\ll{00}(0,\t / 2) \th\ll{00}(0,\tb / 2)
+ \th\ll{01}(0,\t / 2) \th\ll{01}(0,\tb / 2)
\rrdd
\\\\
= 
|-i\t|
\lrdd
\sum\ll{n\ll L, n\ll R}
 \hh(1 + (-1)\uu{n\ll L + n\ll R}) ~\qb\uu{{1\over 4} n\ll L\sqd}
 q\uu{{1\over 4} n\ll R\sqd}
\rrdd
\eee
\bbb
= 
|-i\t|
\lrdd
\sum\ll{n,w}
 \qb\uu{{1\over 4} (n - w)\sqd}
 q\uu{{1\over 4} (n+w)\sqd}
\rrdd
\\\\
= 
|-i\t|\lrdd
\th\ll{00}(0,2\tb)\th\ll{00}(0,2\t)
+ \th\ll{10}(0,2\tb)\th\ll{10}(0,2\t) \rrdd
\\\\
= |-i \t|~z\uu 0\ll 0
\eee
so
\bbb
I\uu 0\ll 0 (\t) \equiv b(\t)~B\uu 0\ll 0(\t) ~z\uu 0\ll 0(\t)
\eee
transforms into itself under $\t\to - {1\over \t}$:
\bbb
I\uu 0\ll 0(- {1\over \t}) = I\uu 0\ll 0 (\t)
\eee
and the invariance under
the T transformation is self-evident due to
level matching.

As for $I\uu 1\ll 0$, the T transformation is clearly still fine.
Under the $S$ transformation 
the components transform as
\bbb
b(\mot) = (-i\t)\uu{-\hh} b(\t)
\\\\
z\uu 1\ll 0(\mot) = {1\over{\sqrt{2}}}~(-i\t)\uu\hh~\th\ll{10}(0,\hh\t)
= {1\over{\sqrt{2}}} (-i\t)\uu\hh z\uu 0\ll 1(\t)
\\\\
B\uu 1\ll 0(\mot) =  \sqrt{2}~\lrdd {{\eta(\tb)}\over{\th\ll{01}(0,\tb)}}
\rrdd\uu\hh
= \sqrt{2} ~B\uu 0\ll 1 (\t)
\eee
with
\bbb
z\uu 0\ll 1(\t) = \sum\ll n q\uu{{1\over 4}(n - \hh)\sqd} = 
\th\ll{10}(0,\hh\t) 
\\\\
= 2 \sum\ll n q\uu{ (n - {1\over 4})\sqd}
\\\\
B\uu 0\ll 1 (\t)= \lrdd {{\eta(\tb)}\over{\th\ll{01}(0,\tb)}}
\rrdd\uu\hh = \qb\uu{-{1\over{24}}}~\qb\uu{+{1\over{16}}}~
\prod\ll{m = 1}\uu\infty (1 - \qb\uu{m - \hh})\uu{-1}
\eee

The factor $B\uu 0\ll 1$ is the partition function for a
set of half-odd-integrally moded left-moving bosonic
oscillators, including
the Casimir energy of $+{1\over{16}}$ from the antiperiodicity of
the real boson.  The factor $z\uu 0\ll 1$ describes a zero
mode sum of a right-moving zero mode only, whose momenta are also
half-odd-integers.  We shall call this the 'twisted sector' of
the T-duality projection.  Indeed, it contains an
infinite number of level matched states, since the $+{1\over{16}}$
coming from the left-moving Casimir energy is balanced by the
$+{1\over{16}}$ coming from the right-moving zero mode state with
momentum $\hh$.  So we define
\bbb
I\uu 0\ll 1 \equiv b~z\uu 0\ll 1~B\uu 0\ll 1
\eee

So far, the natural phases are all correct.  In the last step we will have to
be a bit careful about the phase.

Applying the T transformation, we have
\bbb
b\to \exp{-{{\pi i}\over{12}}} b
\\\\
z\uu 0\ll 1 \to  \exp{{{\pi i}\over 8}} z\uu 1\ll 1
\\\\
B\uu 0\ll 1 \to \exp{-(2\pi i)(- {1\over{24}} + {1\over{16}})}
B\uu 1\ll 1
\eee

where
\bbb
z\uu 1\ll 1 \equiv \sum\ll n \exp{{{\pi i}\over 2}(n\sqd - n)} 
q\uu{{1\over 4}(n - \hh)\sqd}
\\\\
= 2 \sum\ll n (-1)\uu n q\uu{ (n - {1\over 4})\sqd}
\\\\
= \sqrt{2}~\sum\ll n \exp{\pi i ({1\over 4} - {n\over 2})}~
q\uu{{1\over 4} (n - \hh )\sqd}
\\\\
= \sqrt{2}~\th\ll{10}(- {1\over 4},
{\t\over 2})
\\\\
B\uu 1\ll 1 = \qb\uu{-{1\over{24}}}~\qb\uu{+{1\over{16}}}~
\prod\ll{m = 1}\uu\infty (1 + \qb\uu{m - \hh})\uu{-1}
= \lrdd {{\eta(\tb)}\over{\th\ll{00}(0,\tb)}}\rrdd\uu\hh
\eee
So letting $I\ll 1\uu 1 \equiv b z\ll 1\uu 1 B\ll 1\uu 1$,
we have $I\ll 1\uu a (\t + 1) = I\ll a\uu{a+1}(\t)$.
Proving modular invariance now amounts to showing that
$I\uu 1\ll 1$
transforms to itself, with no phase, under $\t\to - {1\over\t}$.
First, check the transformation of $z\ll 1\uu 1$:
\bbb
z\ll 1\uu 1 (- {1\over \t})  = \sqrt{2}~
\th\ll{10} ( - {1\over 4}, - {1\over{2\t}})
\\\\
= \sqrt{2}~
\th\ll{10} ( {1\over {2\t}} \cdot (- {\t\over 2}) ,
- {1\over {2\t}} )
\\\\
=  
\sqrt{2}~\exp{\pi i (- \t / 2)\sqd / (2\t)}~
(- 2 i \t)\uu{\hh} \th\ll{01} (  (- {\t\over 2}) ,
2\t )
\eee
\bbb
= 2 ~ \exp{{\pi i \t \over 8}}~
(- i \t)\uu\hh \th\ll{01} (  (- {\t\over 2}) ,
2\t )
\\\\
= 2~(- i \t)\uu\hh~q\uu{1\over{16}}~
\th\ll{01}(  (- {\t\over 2}) ,
2\t )
\\\\
= 2~(- i \t)\uu{\hh}~
\sum\ll n (-1)\uu n q\uu{(n - {1\over 4})\sqd}
\\\\
= (- i \t)\uu{\hh}~z\ll 1 \uu 1 (\t)
\eee
Next, check that the transformation of $B\uu 1\ll 1$
is trivial:
\bbb
B\uu 1\ll 1 ( - {1\over \t}) = 
B\uu 1\ll 1 (\t)
\eee
so
\bbb
I\ll 1\uu 1(- {1\over \t}) = I\ll 1 \uu 1 (\t)
\eee

\subsection{Fermionic partition functions}
\renewcommand{\fft}{\tilde{F}}

Now we consider partition functions for
left-moving worldsheet fermions
in the type II string on a T-duality
Wilson line background.  T-duality acts
on the fermions $\pst\uu{8,9}$ with
a $-$ sign, as dictated by worldsheet SUSY.
The left-moving fermions therefore break up
into a block of six $\pst\uu{2-7}$ and a block of two
$\pst\uu{8,9}$ which transform differently under $T$.
We will define four partition functions
\bbb
\fft\uu a\ll b,
\eee
with $a$ and $b$ defined mod two, corresponding to
possible cuts by elements of 
the group element $T$, along the two cycles
of the torus.
Each of these four partition functions can be
decomposed further as a sum over four spin structures
for left moving fermions.  For instance,
our definition of the untwisted sector means we can
write
\bbb
\fft\uu a\ll 0\equiv
\hh\lrdd  \fft\uu{a|0}\ll{0|0}
- \fft\uu{a|1}\ll{0|0} - (i\s)\uu a   \fft\uu{a|0}\ll{0|1}
\mp (i \s)\uu a  \fft\uu{a|1}\ll{0|1} \rrdd
\eee
where
\bbb
\fft\uu{a|c}\ll{b|d} \equiv 
(\bfft \uu c\ll d) \uu 3 ~ \bfft \ll{b+d} \uu{a+c}
\eee
and $\s = \pm 1$ is a sign choice describing the $g$-projection 
$g = \pm i$ in the left-moving Ramond sectors.  The possibilities
are $\pm i$ rather than $\pm 1$ because the product
of two identical
spin fields for the $\pst\uu {8,9}$ fermions contains in its
OPE only operators which are odd, rather than even,
under $g$.  This can be seen easily through bosonization.

The $\bfft\ll b\uu a$, defined as traces of
$\qb\uu{\tilde{L}\ll 0}$ in the sector twisted by $b$,
appropriately GSO projected and summed over R and NS states,
sometimes have a nontrivial eigenvalue of $g$ even in
the NS$_+$ sector.  In the NS sector twisted by $g\uu b$,
the ground state gets a phase $i\uu b$ under $g$ due to
fermion zero modes.  Therefore the contribution $\fft\uu{a|0}
\ll{b|0}$ to $\fft\uu a\ll b$ 
has a coefficient $i\uu{ab}$.  Combined with the 
requirement that the $\fft \uu a\ll b$ transform into
one another under modular transformations up to a phase,
this uniquely fixes the $\fft\uu a\ll b$.  We have
\bbb
\fft\uu a\ll b \equiv \hh \sum\ll{c,d}
\o\uu {a|c}\ll{b|d} 
(\bfft\uu c\ll d)\uu 3 \bfft\uu {a+c}\ll{b+d}
\eee
with
$\o\uu {a|0}\ll{b|0} = i\uu{ab}$ and $|\o\uu {a|c}\ll{b|d}| = 1$.
The values of $\fft\uu a\ll b$ with $a$ and $b$ running from
0 to 1 are
\bbb
\fft\uu 0\ll 0 (\tb) = \hh \lrdd \fft\uu {0|0} 
\ll{0|0} (\tb)
- \fft\uu{0|1}\ll{0|0}(\tb)
- \fft\uu{0|0}\ll{0|1}(\tb)
  \mp \fft\uu{0|1}\ll{0|1} (\tb)
\rrdd
\\\\
\fft\uu 1\ll 0 (\tb) =\hh \lrdd \fft\uu {1|0} 
\ll{0|0} (\tb)
- 
\fft\uu{1|1}\ll{0|0}(\tb) 
-i \fft\uu{1|0}\ll{0|1}(\tb)
 \mp i \s \fft\uu{1|1}\ll{0|1}(\tb) \rrdd
\\\\
\fft\uu 0\ll 1 (\tb) = \hh \lrdd \fft\uu {0|0} 
\ll{1|0} (\tb)
+ i \s
\fft\uu{0|1}\ll{1|0}(\tb) 
- \fft\uu{0|0}\ll{1|1}(\tb)
 \pm i \s \fft\uu{0|1}\ll{1|1}(\tb) \rrdd
\\\\
\fft\uu 1\ll 1 (\tb) =
{i\over 2} \lrdd \fft\uu{1|0}\ll{1|0} - i \s \fft\uu{1|1}
\ll{1|0}
- \s \fft\uu{1|0}\ll{1|1} \pm i \fft\uu{1|1}\ll{1|1} \rrdd
\eee
The phases have a periodicity mod 2
which defines the rest: 
\bbb
\o\uu{a+2p|c}\ll{b+2q|d}
 = (-1)\uu{p(b+d)+q(a+c)} \o\uu{a|c}\ll{b|d}
\eee


\end{document}